\documentclass[11pt]{article}
\pdfoutput=1  
\usepackage{jcappubmod}
\usepackage{graphicx}
\usepackage{color}
\usepackage{amssymb}
\usepackage{amsmath}
\usepackage{mathabx}
\usepackage{stmaryrd}
\usepackage{multirow}
\usepackage{multicol}
\usepackage{amscd}
\usepackage{mfpic}
\usepackage{verbatim}
\usepackage{hyperref}
\usepackage{pstricks,pst-node,pst-text,pst-3d}
\usepackage{ftnxtra}
\usepackage{fnpos} 

\definecolor{cream}{rgb}{.97, .95, .88}
\definecolor{darkcream}{rgb}{1., .88, .5}
\definecolor{lightpink}{rgb}{0.98, 0.88, 0.87}
\definecolor{lightwhite}{rgb}{1., 0.98, 0.95}
\definecolor{lightsalmon}{rgb}{1., 0.95, 0.90}
\definecolor{lightviolet}{rgb}{0.9, 0.8, 0.9}
\definecolor{lightgray}{rgb}{.96, .96, .96}  
\definecolor{lgray}{rgb}{.75, .75, .75}
\definecolor{LemonChiffon}{rgb}{0.95, 1., 0.7}
\definecolor{lightolivegreen}{rgb}{0.84, 0.89, 0.25}
\definecolor{lightgreen}{rgb}{.664, 1., .52}
\definecolor{llgreen}{rgb}{.900, .983, .960}
\definecolor{tristle}{rgb}{0.87, 0.67, 0.77} 
\definecolor{pink}{rgb}{0.95, 0.45, 0.75}
\definecolor{magenta}{rgb}{1., 0, 1.}
\definecolor{violet}{rgb}{0.9, 0.20, 0.85}
\definecolor{darkolivegreen}{rgb}{0.55, 0.65, 0.35}
\definecolor{maroon}{rgb}{0.7, 0.26, 0.56}
\definecolor{lightmaroon}{rgb}{0.85, 0.38, 0.58}
\definecolor{darkmaroon}{rgb}{0.604, 0.169, 0.451}
\definecolor{ddarkmaroon}{rgb}{0.2, 0.03125, 0.150}
\definecolor{mediumorchid}{rgb}{0.8, 0.33, 0.83}
\definecolor{mediumorchidd}{rgb}{1., 0.33, 0.63}
\definecolor{darkgreen}{rgb}{0.1, 0.6, 0.13}
\definecolor{lightyellow}{rgb}{1., 1., 0.82}
\definecolor{turquoise}{rgb}{0.042, 0.586, 0.512}
\definecolor{turquoisel}{rgb}{0.66, 0.94, 0.83}
\definecolor{darkturquoise}{rgb}{0.21, 0.55, 0.50}
\definecolor{coral}{rgb}{1., 0.6, 0.21}
\definecolor{lightorange}{rgb}{1., 0.88, 0.75}
\definecolor{orangered}{rgb}{1., 0.5, 0.}
\definecolor{orange}{rgb}{1., 0.65, 0.1}
\definecolor{orangel}{rgb}{1., .85, .3}
\definecolor{darkorange}{rgb}{0.875, 0.4, 0.204}
\definecolor{ddarkorange}{rgb}{.675, .218, .05}
\definecolor{bluesky}{rgb}{0.48, 0.53, 1.}
\definecolor{gold}{rgb}{1., 0.85, 0.25}
\definecolor{goldd}{rgb}{0.95, 0.75, 0.05}
\definecolor{darkviolet}{rgb}{0.54, 0.04, 0.84}
\definecolor{ddarkviolet}{rgb}{.382, .063, .657}
\definecolor{lightblue}{rgb}{0.30, 0.86, 0.89}
\definecolor{LightBlue}{rgb}{0.68, 0.85, 0.9}
\definecolor{lblue}{rgb}{0.78, 0.90, 0.95}
\definecolor{darkblue}{rgb}{.105, .308, .707}
\definecolor{lightmaroon}{rgb}{0.85, 0.38, 0.58}
\definecolor{darkmaroon}{rgb}{0.604, 0.169, 0.451}
\definecolor{darkpink}{rgb}{0.879, 0.020, 0.766}
\definecolor{ddarkpink}{rgb}{0.738, 0.195, 0.406}
\definecolor{grey}{rgb}{0.717, 0.717, 0.717}
\definecolor{lightgrey}{rgb}{0.800, 0.800, 0.800}
\definecolor{brown}{rgb}{0.740, 0.323, 0.182}
\definecolor{redbrown}{rgb}{.575, .158, .05}
\definecolor{darkbrown}{rgb}{0.34, 0.25, 0.05}
\definecolor{orangebrown}{rgb}{0.433, 0.262, 0.06}
\definecolor{pinkl}{rgb}{1., 0.788, 0.918}
\definecolor{salmon}{rgb}{1., 0.66, 0.5}
\definecolor{lightbrown}{rgb}{0.703, 0.508, 0.121}

\def\etal{{\it et al.}}

\def\Name#1#2 {{#2} {#1, }}
\def\Journal#1#2#3#4{{#3}, {#1}, {\bf #2}, #4}
\def\cir#1{{\GCN} #1}
\def\rep#1{{\GCR} #1}
\def\AA{\em A.\& A.}

\def\AAS{\em A.\& A. Suppl.}

\def\AIP{\em AIP Conf.Proc.}

\def\APJ{\em ApJ.}
\def\APJL{\em ApJ.Lett.}
\def\APJS{\em ApJ.Suppl.}

\def\ARA{\em Annu.Rev.A\&A}

\def\APS{\em APS }

\def\CQG{\em Class.Quant.Grav.}

\def\GCN{\em GCN Circ.}
\def\GCR{\em GCN Rep.}

\def\JCA{\em J. Cosmol. Astrop. Phys.}

\def\MRA{\em MNRAS}

\def\NAT{\em Nature}

\def\NATP{\em Nature Phys.}

\def\PRD{{\em Phys. Rev.} D}
\def\PRL{\em Phys. Rev. Lett.}

\def\RPP{\em Rept. Prog. Phys.}
\def\SCI{\em Science}

\def\SSC{\em Space Sci.}

\def\be{\begin{equation}}
\def\ee{\end{equation}}
\def\bea{\begin{eqnarray}}
\def\eea{\end{eqnarray}}
\def\bes{\begin{equation*}}
\def\ees{\end{equation*}}
\def\beas{\begin{eqnarray*}}
\def\eeas{\end{eqnarray*}}

\begin{document}
\title{Binary Neutron Star (BNS) merger: What we learned from relativistic ejecta of GW/GRB~170817A}

\author[a,b]{Houri~Ziaeepour}
\affiliation [a]{Institut UTINAM, CNRS UMR 6213, Observatoire de Besan\c{c}on, Universit\'e de Franche Compt\'e, 41 bis ave. de l'Observatoire, BP 1615, 25010 Besan\c{c}on, France}
\affiliation [b]{Mullard Space Science Laboratory, University College London, Holmbury St. Mary, 
GU5 6NT, Dorking, UK}

\emailAdd{houriziaeepour@gmail.com}


\abstract
{Gravitational waves from coalescence of a Binary Neutron Star (BNS) and its accompagning short 
Gamma-Ray Burst GW/GRB~170817A confirmed the presumed origin of these puzzeling transients and 
opened up the way for relating properties of short GRBs to those of their progenitor 
stars and their surroundings. Here we review an extensive analysis of the prompt gamma-ray and late 
afterglows of this event. We show that a fraction of polar ejecta from the merger had been 
accelerated to ultra-relativistic speeds. This structured jet had an initial Lorentz factor of about 
$260$ in our direction - $\mathcal{O}(10^\circ)$ from the jet's axis - and was a few orders of 
magnitude less dense than in typical short GRBs. At the time of arrival to circum-burst material 
the ultra-relativistic jet had a close to Gaussian profile and a Lorentz factor $\gtrsim 130$ in 
its core. It had retained in some extent its internal collimation and coherence, but had extended 
laterally to create mildly relativistic lobes - a {\it cocoon}. External shocks on the far from 
center inhomogeneous circum-burst material and low density of colliding shells generated slow rising 
afterglows. The circum-burst material was somehow correlated with the merger and it is possible 
that it contained recently ejected material from glitching, which had resumed due to the deformation 
of neutron stars crust by tidal forces in the latest stages of inspiral but well before their merger. 
By comparing these findings with the results of relativistic MHD simulations and observed 
gravitational waves we conclude that progenitor neutron stars were old, had close masses and highly 
reduced magnetic fields. In addition, they probably had oppositely directed spins due to the 
encounter and gravitational interaction with other stars.

}

\maketitle

\section {Introduction} \label{sec:intro}
As a light dominated species we are usually more confident about nature of objects when we can 
see them in our very restricted electromagnetic band of $380$~nm to $740$~nm - {\it the visible band}. 
For this reason since the discovery of Gamma-Ray Bursts (GRBs) in 1960's efforts for understanding 
these enigmatic transients have relied on detecting their counterparts in optical and other 
electromagnetic bands with our extended {\it eyes} - ground based and space telescopes. Observation 
of X-ray counterpart of GRB~970228~\cite{grb970228bepposax,grb970228bepposaxx} by 
BeppoSAX~\cite{bepposax} led to its detection in optical by the William Herschel 
Telescope~\cite{grb970228opt}. For the first time these observations proved that GRBs are 
extragalactic and in contrast to prediction of some models, have a broad band of electromagnetic 
emissions. Since then observation of thousands of GRBs and their afterglows in X-ray, for roughly 
all confirmed GRB detected by the Neil Gehrels Swift~\cite{swift} and other spatial Gamma-ray 
observatories such as BATSE~\cite{batse} and Fermi~\cite{fermi}, and in optical/IR and radio for 
large fraction of them, have clarified many aspects of origin and physical mechanisms involved in 
their production. Notably, discovery of supernova type Ib and Ic associated to long GRBs confirmed 
collapsar hypothesis and proved that they are produced by ultra-relativistic jets ejected from 
exploding massive stars - also called {\it collapsars} - at the end of their life. However, no direct 
evidence of the origin of short GRBs was available until the breakthrough observation of 
gravitational waves from merger of a binary neutron stars by Ligo~\cite{ligo} and Virgo~\cite{virgo}. 

In many respects the short GRB~170817A associated to the first detection of gravitation waves 
from a Binary Neutron Star (BNS) 
merger~\cite{gw170817ligo,gw170817ligoprogen,gw170817multimess,gw170817fermimulti} was unusual: 
\begin{itemize}
\item The prompt gamma-ray was intrinsically faint and somehow softer than any other short GRB with 
known redshift; 
\item Its X-ray afterglow was not detected - despite several attempts as early as 
$\sim T+1.6$~days~\cite{gw170817swiftnustar,gw170817cxc2day}, where $T$ is the Fermi trigger time;
\item It was finally detected at $\sim T+10$~days~\cite{gw170817xray}, but surprisingly rather than 
fainting, its flux increased and peaked at $\sim T+110$~days; 
\item The same behaviour was observed in radio bands; 
\item By contrast, the kilonova emission~\cite{kilonovarev,gw170817rprocess,gw170817optkilonovath} 
in UV was brighter than expected~\cite{gw170817bluekilonova}.
\end{itemize}
Initially, the simplest explanation seemed to be an off-axis view of an otherwise ordinary 
short GRB~\cite{grboffaxis,grboffaxprobab} with a uniform (top hat) or structured 
ultra-relativistic jet~\cite{structuredjet,gw170817latexraystructjet}. Alternatively, the burst 
might have been formed by a mildly relativistic magnetized cocoon, i.e. an outflow with a Lorentz 
factor $\sim 2-3$ at its breakout~\cite{grbcocoon} from the envelop of merged 
stars~\cite{gw170817cocoon,gw170817cocoon0,gw170817earlyradio,gw170817lateradio,gw170817cocoonsimul,gw170817cocoonevid}. 
However, further observations and detection of the decline of flux in all 3 observed energy bands, 
i.e. radio~\cite{gw170817lateradio,lateradio,gw170817latexoptradio1,gw170817lateradio300}, optical~\cite{gw170817lateopt,gw170817lateopt160,gw170817lateopthstir}, 
and X-ray~\cite{gw170817latexary,gw170817chandraxray260,gw170817chandraxray260a,gw170817latexraydecline,gw170817xraycxc260,gw170817chandraxray358} 
after $\gtrsim T+200$ days was much earlier than the prediction of a fully off-axis jet, or a 
cocoon/jet breakout. In addition, detailed modelling of the prompt gamma-ray~\cite{hourigw170817} 
showed that the most plausible initial Lorentz factor for the jet was $\mathcal{O}(100)$ but 
significantly less than those of typical short GRBs~\cite{hourigrbmag,hourigw170817}. Its density 
along our line of sight was also much less than other short GRBs. 

Gradually it became clear that an additional source of X-ray~\cite{hourigw170817ag} or presence 
of a highly relativistic component with a Lorentz factor of $\mathcal{O}(10)-\mathcal{O}(100)$ 
in the outflow at late times is inevitable~\cite{gw170817latexary,gw170817lateopthstir,hourigw170817lateagjet}. 
This meant that at least a fraction of the initial ultra-relativistic jet had survived internal 
shocks and energy dissipation up to long distances. These conclusions made from properties of 
afterglows are consistent with those obtained from analysis of the prompt gamma-ray emission 
alone. On the other hand, detection of the superluminal motion of the radio 
afterglow~\cite{gw170817lateradiosuprlum,gw170817lateradiosuprlum0} proved off-axis view 
of the late time jet, and simulations of emissions from such a system led 
to estimation of a viewing angle $\theta_v \sim 10^\circ-20^\circ$ for its 
source~\cite{gw170817latexoptradio1,gw170817lateradiosuprlum,hourigw170817lateagjet}.

In this review we first briefly summarize multi-probe multi-wavelength observations of 
GW/GRB~170817A in Sec. \ref{sec:obs} and compare some of properties of this transient with other 
short GRBs. Then, in Sec. \ref{sec:anal} we use a phenomenological model of relativistic shocks 
and synchrotron/self-Compton emission~\cite{hourigrb,hourigrbmag} (reviewed in~\cite{hourigrbrev}) to 
analyze both prompt gamma-ray and afterglows of this GRB. Interpretation of models and what they 
teach us about ejecta from the merger and its environment are discussed in Sec. \ref{sec:interpret}. 
In Sec. \ref{sec:progenitor} we use predictions of GRMHD simulations and physics of neutron stars 
and compare them with conclusions obtained from multi-wavelength observations of GW/GRB~170817A and 
their analysis to see what we can learn about progenitor neutron stars, their life history, 
physical properties, and environment just before their coalescence. Finally, in 
Sec. \ref{sec:prespect} as outline we give a qualitative description of this transient and briefly 
discuss prospectives for deeper understanding of compact objects with increasing data and new probes. 

We should remind that this review is concentrated on the GRB component of the merger and does not 
address other components, namely the ejected disk/torus and its kilonova emission and the merger 
remnant. They are very important subjects by their own and need separate analyses.

\section {Review of multi-probe, multi-wavelength observations of GW/GRB~170817A} \label{sec:obs}
GW~170817A was the first gravitation wave source detected in a panoramic range of electromagnetic 
energy bands. In this section we briefly review its observations and their findings.

\subsection {Gravitational waves (GW)} \label{sec:gw}
At 12:41:04 UTC on 17 August 2017 the Advanced Ligo and Advanced Virgo gravitational wave detectors 
registered a low amplitude signal lasting for $\sim 100$~sec, starting at $24$~Hz 
frequency~\cite{gw170817ligo}. These are hallmarks of gravitational wave emission from inspiral of 
low mass $M \sim \mathcal {O}(1) M_{\odot}$ compact binary objects, where $M_\odot$ is the solar mass. 
Further analysis of gravitational wave data confirmed a chirp mass 
$\mathcal{M} \equiv (m_1 m_2)^{3/5}/(m_1 + m_2)^{1/5}\approx 1.1977~M_\odot$, a total mass 
$M \approx 2.74~M_\odot$, and individual masses of $m_1 \approx (1.36~-~1.6)~M_\odot$ and 
$m_2 \approx (1.17~-~1.36)~M_\odot$~\cite{gw170817ligoprogen}. These masses are in neutron star mass 
range and led to conclusion that a BNS merger was the source of the observed GW signal. 

Due to low resolution of Ligo-Virgo detectors at frequencies $\gtrsim 500$~Hz - necessary for 
analysing ringdown signal - estimation of radius, tidal deformability, and Equation of States (EoS) 
of progenitors are model dependent. Two different sets of assumptions 
in~\cite{gw170817state,gw170817state0,gw170817ligostate} led to following constraints: Radius of 
the progenitors $R_1 \sim (10.8~-~11.9) \pm 2$~km and $R_2 \sim (10.7~-~11.9) \pm 2$~km; Tidal 
deformability parameters $\Lambda_1 < \Lambda_2$, $\Lambda_1 < 500$ and $\Lambda_1 < 1000$; 
Stiff EoS such as H4 and MPA1 disfavored.

\subsection{Prompt gamma-ray} \label{sec:gamma}
GRB~170817A was detected by the Fermi-GBM~\cite{gw170817fermi} and the 
Integral-IBIS~\cite{gw170817integral} detectors at about $1.7$~sec after the end of inspiral stage 
of gravitational waves. It lasted for about $2$~sec, had an integrated fluence of 
$(2.8 \pm 0.2) \times 10^{-7}$ erg cm$^{-2}$ in the 10~keV to 1~MeV Fermi-GBM band and 
$(1.4 \pm 0.4 \pm 0.6) \times 10^{-7}$ erg~cm$^{-2}$ in the Integral-IBIS 75~keV to 2~MeV 
band. The peak energy was $E_{peak} = 229 \pm 78$ keV. Unfortunately at the time of prompt gamma-ray 
emission the source was not in the field of view of the Swift-BAT. For this reason no early 
follow-up data from $\lesssim T+1.6$~days is available, except for an upper limit of $> 4-$sigma 
on any excess from background in 10~keV to 10~MeV band from Konus-Wind satellite~\cite{gw170817konus}.

Despite lack of early follow up, detection of the gravitational waves by both Ligo and Virgo resulted 
to an sky-localization area much smaller than previous GW events. This helped confirmation of 
coincidence between GRB~170817A and GW~170817 and follow up of the event in low energy 
electromagnetic bands~\cite{gw170817ligoprogen}. The host galaxy of the transient was identified to be 
NGC 4993~\cite{gw170817optsss1,gw170817optdes} at $z = 0.0095$, that is at a distance 
of $\sim 40$~Mpc\footnote{Here we use vanilla $\Lambda$CDM cosmology with 
$H_0 = 70$~km sec$^{-1}$~Mpc$^{-1}$, $\Omega_m = 0.3$ and $\Omega_\Lambda = 0.7$.}, making GRB~170817A 
the closest GRB with known distance so far, see e.g.~\cite{sgrbrev} for a review of properties 
of short GRBs and their hosts.

\subsection {X-ray afterglow} \label{sec:xray}
The earliest observation of GW/GRB~170817A in X-ray was at about $T+0.6~\text{days} = T+51840$~sec 
by the Swift-XRT~\cite{gw170817swiftnustar}. There is also an upper limit of 
$\sim 10^{-12}$~erg~sec$^{-1}$~cm$^{-2}$ in 0.3-10~keV band at around $T+0.2$~days on any excess of 
X-ray flux obtained from observations of the Swift-XRT in the sky area calculated from gravitational 
wave signal. Upper limit on the early X-ray afterglow from Chandra observations is 
$3.5 \times 10^{-15}$ erg~sec$^{-1}$~cm$^{-2}$ at $\sim T+2$~days~\cite{gw170817cxc2day}. Although 
this limit is lower than flux of any previous short GRB with known redshift at similar 
epoch~\cite{gw170817grbcomp}, it is higher than early flux of many short GRBs without redshift. 
Therefore, in an observational sense it is not very restrictive.

A X-ray counterpart was finally found by Chandra observatory~\cite{gw170817cxc,gw170817xray} at 
$\sim T+9$ days with a flux of $\sim 2.7 \times 10^{-15}$ erg~sec$^{-1}$~cm$^{-2}$ in 0.3-8~keV. 
Moreover, further observations~\cite{gw170817latexary} showed that the afterglow was gradually 
brightening. Late time brightening in X-ray and optical is claimed in a few other 
short~\cite{grb130603bxray} and long~\cite{grb060712latex,grb060807latex} GRBs. However, 
in contrast to prediction of off-axis models that brightening could last several hundreds of days, 
X-ray light curve peaked somewhere in the interval of 
$\sim 110 - 134$~days~\cite{gw170817latexraydecline} and began to decline 
afterward~\cite{gw170817chandraxray260,gw170817chandraxray260a,gw170817xraycxc260,gw170817chandraxray358}.

\subsection {Optical/IR afterglow}  \label{sec:opt}
Despite low spatial resolution of Ligo-Virgo and Fermi-GBM, follow up of GW/GRB~170817A by a 
plethora of ground and space based telescopes~\cite{gw170817multimess,gw170817fermimulti} allowed 
to find optical/IR counterpart of the transient, known also as 
AT~2017~gfo~\cite{gw170817optkilonova,gw170817optdes}, SSS17a~\cite{gw170817optsss1,gw170817optsss}, 
and DLT17ck~\cite{gw170817optdlt}. The earliest detection of the optical counterpart was at 
$\sim 41$~ksec~$\sim T+0.5$~days and further observations were performed at $\sim T+1.6$~days 
onward~\cite{gw170817optdes}. Its magnitude was: $r,~i \sim 17$, 
$u \sim 19$~\cite{gw170817optdes,gw170817optkilonova} and 
$B \sim 20$~\cite{gw170817optsss1,gw170817optsss}. Spectroscopy data showed that optical/IR flux 
was dominated by a kilonova emission~\cite{gw170817bluekilonova,gw170817optkilonovath}. However, 
the UV emission seemed to be too bright. The initial conclusion was that the afterglow of the GRB 
might have had contributed, otherwise the mass of slow kilonova ejecta had to be larger than 
$\sim 0.01-0.03 M_\odot$ predicted by kilonova models~\cite{gw170817optkilonovath,gw170817bluekilonova}.

Later observations~\cite{gw170817lateopt,gw170817latexary,gw170817lateopt160,gw170817latexoptradio1,gw170817lateopthstir} 
showed that as expected, the kilonova/afterglow emission had rapidly declined. Unfortunately between 
$\sim T+16$~days and $110$~days there was no follow up observation in optical/IR bands. Nonetheless, 
later observations showed that the decline of the source continued with somehow shallower slope 
until $\sim T+200$~days and then became steeper. The shallower slope in $\sim T+100~-~T+200$ can be 
interpreted as when the decay of isotopes and cooling of kilonova 
ejecta~\cite{gw170817latexary,gw170817lateopt,kilonovarev} had reduced its optical/IR flux and 
contribution of GRB afterglow in visible bands became dominant.

\subsection {Radio afterglow}  \label{sec:radio}
The counterpart of GW/GRB~170817 was also observed in GHz radio band relatively early, that is at 
$\sim T+16$~days onward ~\cite{gw170817earlyradio,gw170817earlyradio1}. But, earlier observations 
at $\sim T+2.4$~days didn't find any excess in the direction of the 
source~\cite{gw170817earlyradio1}. Thus, similar to the X-ray emission the afterglow had been 
brightening. This conclusion was indeed confirmed by further 
observations~\cite{gw170817latexary,gw170817lateradio} and was considered as a confirmation of 
off-axis view of a relativistic jet~\cite{gw170817latexary} or a mildly relativistic large opening 
angle outflow (a cocoon)~\cite{gw170817lateradio}. The peak and turnover of the light curve was 
observed at $\sim T+150$~days~\cite{lateradio} and confirmed by later 
observations~\cite{gw170817latexoptradio1,gw170817lateradiosuprlum,gw170817lateradio300}. 
They ruled out a highly off-axis/side view of a relativistic jet or break out of a 
mildly relativistic cocoon as the origin of this weak GRB. 

Overall, these observations showed that at late times the jet had to include a relativistic 
component~\cite{gw170817latexary,gw170817latexoptradio1,gw170817lateradiosuprlum,gw170817lateradio300,gw170817lateopthstir,hourigw170817lateagjet}. 
In addition, the detection of superluminal motion of the radio source with an estimated apparent 
velocity of $\beta_{app} = 4.1 \pm 0.5$~\cite{gw170817lateradiosuprlum,gw170817lateradiosuprlum0} 
confirmed its oblique view and allowed to obtain a lower limit for source's Lorentz 
factor. This information along with simulation of the jet and consistency with other observations 
led to estimation of viewing angle $\theta_v \sim 10^\circ - 20^\circ$ with respect to symmetry axis 
of the jet~\cite{gw170817latexary,gw170817latexoptradio1,gw170817lateradiosuprlum,hourigw170817lateagjet}. 
This off-axis angle is consistent with estimation of the orbit inclination 
$18^\circ \lesssim \theta_{in} \lesssim 27^\circ$~\cite{gw170817decline} using only gravitational wave data.

\subsection{Comparison with other short GRB-kilonova events} \label{sec:obscomp}
\begin{figure}
\begin{center}
\begin{tabular}{p{5cm}p{5cm}}
\includegraphics[width=7cm,angle=90]{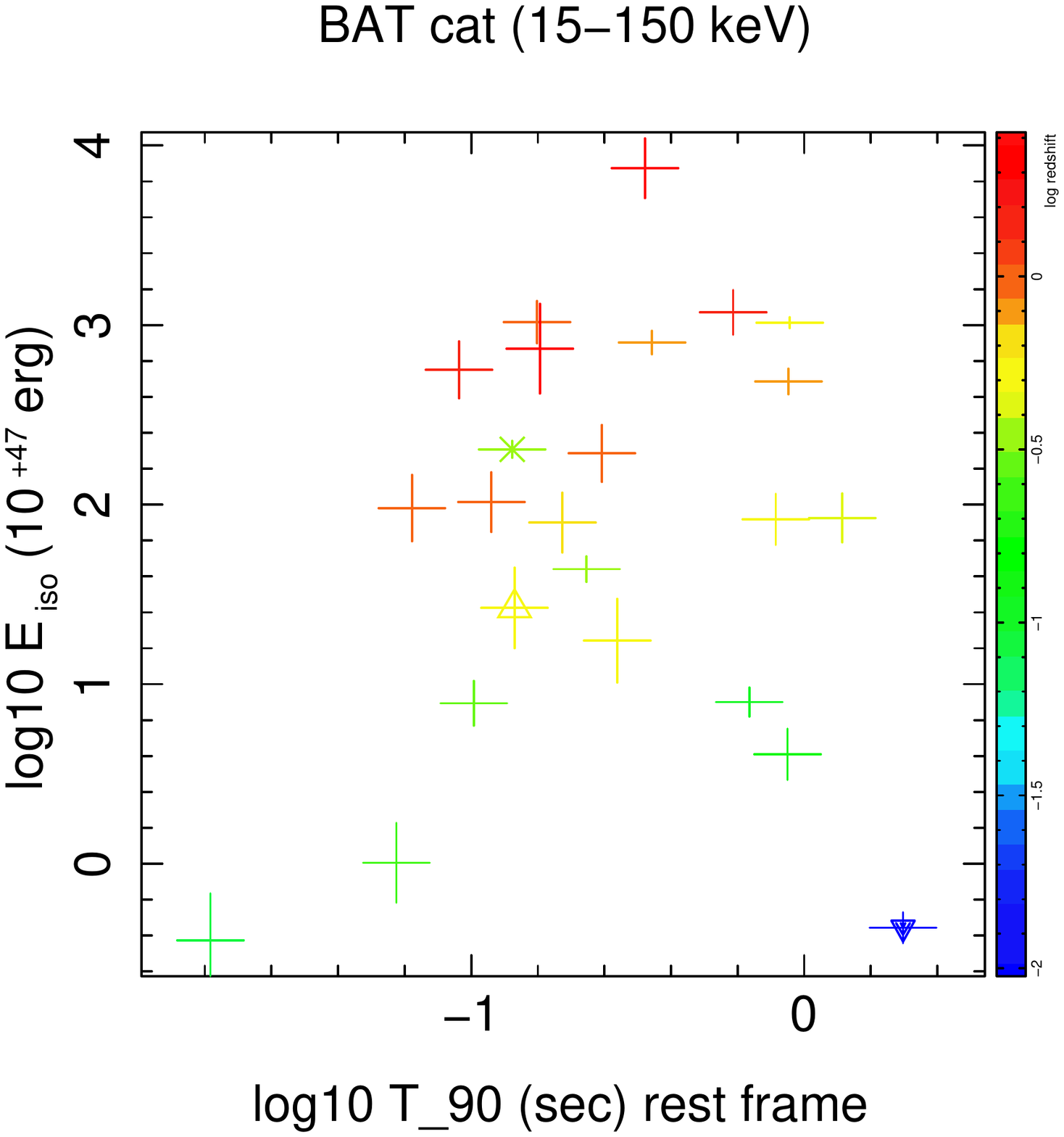} & 
\hspace{-1.5cm} \includegraphics[width=7cm,angle=90]{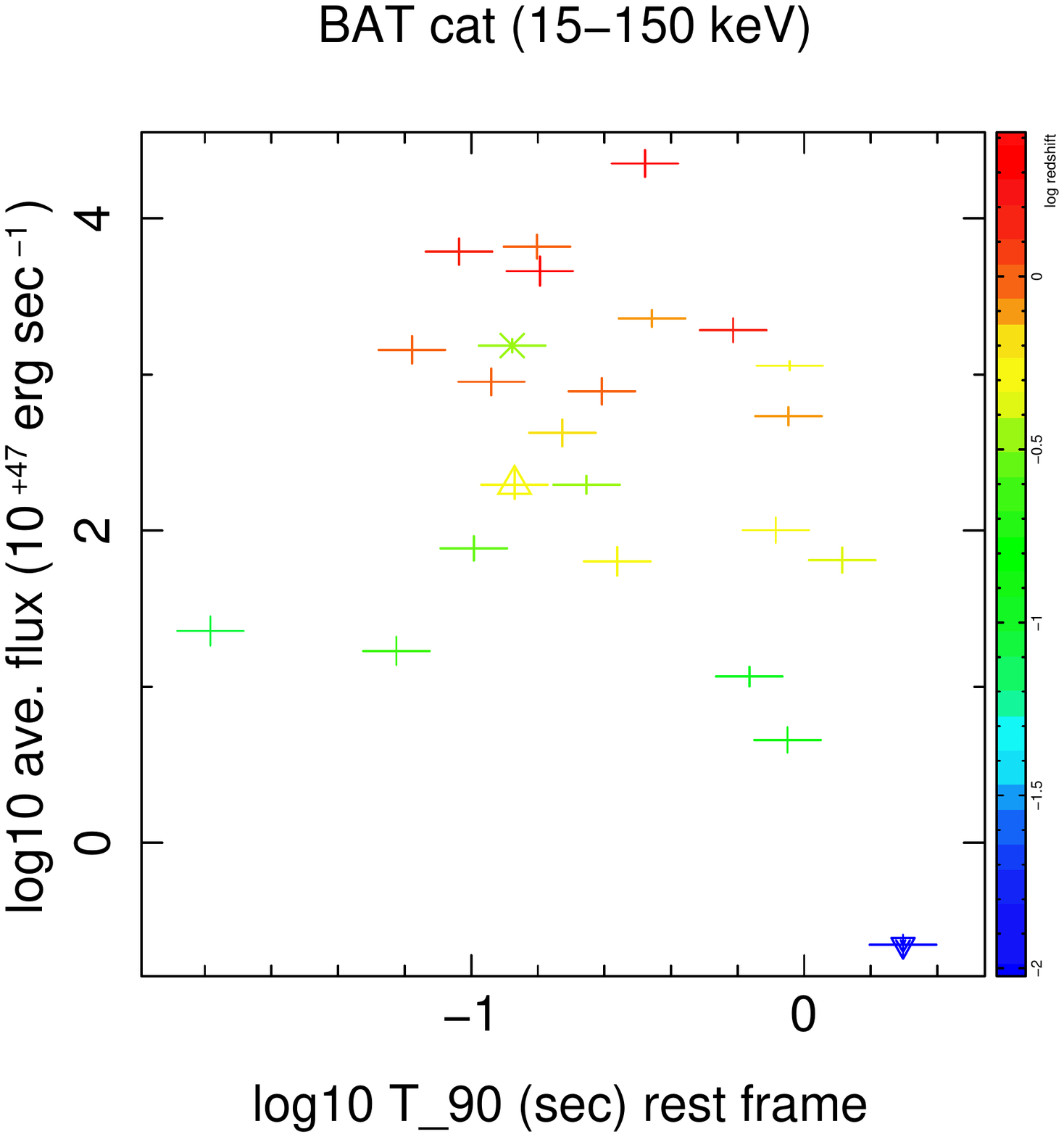}
\end{tabular}
\caption{Left: $E_{iso}$ of short GRB's with known redshift in the Swift-BAT $15-150$~keV energy band. 
Right: Average flux of the same data. The data is taken from the Swift GRB on-line database 
\href{https://swift.gsfc.nasa.gov/archive/grb\_table/}{https://swift.gsfc.nasa.gov/archive/grb\_table/} 
using as selection criteria $T_{90} \leqslant 2$ sec. Redshift is color coded. As GRB~170817A was not 
in the FoV of the Swift-BAT, we have used fluence measured in the Fermi-GBM 10~keV-2~MeV band. Thus, 
$E_{iso}$ and average flux of GRB~170817A shown here are upper limits and shown with an inverse 
triangle as the symbol of upper limit. Star symbol presents kilonova/GRB~130603B and up-right 
triangle is GRB~160624A at $z=0.483$, the only GRB with known redshift since 01 September 2015, 
considered as the beginning of the Advanced LIGO operation, which its GW could be apriori observed 
if it was at a lower redshift. \label{fig:allsgrb}}
\end{center}
\end{figure}

With an isotropic luminosity $E_{iso}\sim 5 \times 10^{46}$ erg in 10 keV to 1 MeV band, GRB~170817A is 
intrinsically the faintest short burst with known redshift, see Fig. \ref{fig:allsgrb}. Moreover, the 
peak energy of the burst was close to the lowest peak energy of short bursts observed by Fermi-GBM, 
see Fig. 31 in~\cite{grbgbmspect}. Therefore, it is not a surprise that its afterglows, specially 
in X-ray, were the faintest among short bursts with known redshift, see e.g.~\cite{gw170817grbcomp}. 
However, it is not sure that GW/GRB~170817 could be classified as X-ray dark at early times, see 
Fig. \ref {fig:darkgrb} and Sec. \ref{sec:offdark} for more discussion. Some short GRB's such as 
GRB~070724A~\cite{grb070724a,grb070724aopt,grb070724arpoc}, GRB~111020A~\cite{grb111020a}, 
GRB~130912A~\cite{grb130912a,grb130912aopt}, and GRB~160821B~\cite{grb160821b,grb160821bir} had 
similar or smaller fluxes at $\sim T+2$~days after trigger. Notably, GRB~111020A is an interesting 
case because its host galaxy is most probably at redshift 0.02~\cite{grb111020aredshift}. It had a 
flux of $\sim 1.2 \times 10^{-11}$ erg~sec$^{-1}$~cm$^{-2}$ in 0.3-10~keV at $\sim T+100$~sec, which 
had decayed to $\sim 8 \times 10^{-15}$ erg~sec$^{-1}$~cm$^{-2}$ at $\sim T+9$ days. 

The outline of this section is the importance of early discovery of electromagnetic (EM) counterpart 
of gravitational wave events for the identification and study of their sources. Specifically, 
it is crucial to improve angular resolution of gravitational wave detectors by multiplying their 
number, and gamma-ray telescopes by increasing their surface and resolution. This will help to reduce 
sky area to be searched for afterglows in lower energies.

\section {Analysis and modeling of GW/GRB~170817A data} \label{sec:anal}
Observations of thousands of gamma-ray bursts and their afterglows by Swift, Fermi, Konus-wind, and 
other high energy satellites and ground based robotic telescopes have demonstrated that the main 
component in the prompt gamma-rays and their afterglows in lower energies is synchrotron emission 
generated in relativistic or mildly relativistic shocks. In this section we first briefly review a phenomenological 
shock and synchrotron emission formalism developed in~\cite{hourigrb,hourigrbmag}. Then, we use 
it to analyse both prompt gamma-ray and afterglows of GW/GRB~170817A. A more extensive review of the model can be 
found in~\cite{hourigrbrev}. The advantage of this formalism, despite its simplicity, is that it can 
be applied to both intern and external shocks, and thereby make it possible to construct a consistent and 
overall picture about: relativistic ejecta from compact objects such as collapsars, BNS and NS-BH mergers; clarify 
relations between properties of jets and their progenitors; and characteristics of the environment 
around progenitors, which influence GRB afterglows.

\subsection {Phenomenological formulation of relativistic shocks and synchrotron/self-Compton emission} \label{sec:model}
The phenomenological model of~\cite{hourigrb,hourigrbmag} assumes that GRB emissions are 
synchrotron/self-Compton produced by accelerated charged leptons in a dynamically active region 
in the head front - {\it wake} - of shocks between density shells inside a relativistic jet for 
prompt and with surrounding material for afterglows in lower energies. In addition to the magnetic 
field generated by Fermi processes in the active region, an external magnetic field precessing with 
respect to the jet axis may contribute in the production of synchrotron emission. Origin of this field 
is not relevant for this simple model. It can be the magnetic field of central object or Poynting flow 
imprinted into the jet/outflow.

An essential aspect of this model, which distinguishes it from other phenomenological GRB 
formulations, is the evolution of parameters with time. Moreover, simulation of each burst consists 
of a few time intervals - {\it regimes} - each corresponding to an evolution rule (model) for 
phenomenological quantities such as fraction of kinetic energy transferred to fields and its 
variation, variation of the thickness of synchrotron/self-Compton emitting {\it active} region, etc. 
Division of simulated bursts to these intervals allows to change parameters and phenomenological 
evolution rules which are kept constant during one time interval. Physical motivation for such 
fine-tuning is the fact that GRB producing shocks are in a highly non-equilibrium and fast varying 
state. In fact, multiple variation of the slope of afterglows light curves, which presumably are 
produced by external shocks on the ISM or circumburst material is an evidence that they are not 
completely uniform and their anisotropies affect the emission. Definition of parameters of the model 
and phenomenological expressions used for the evolution of active region, which in the framework of 
the model cannot be calculated from first principles are given in Appendix \ref{app:paramdef}. 

An important issue, specially when considering the best models for a specific burst, is the fact that 
parameters of model are not completely independent from each others. For instance, fractions of 
kinetic energy transferred to induced electric and magnetic fields should depend on the strength 
of the shock, which in turn is determined by the density difference of colliding shells and their 
relative Lorentz factor. But there is no simple first principle way to incorporate these dependencies 
to the model. 

Originally, the goal of the shock formalism developed in ~\cite{hourigrb,hourigrbmag} was modelling 
prompt and early afterglow emissions, which are presumably generated by a fast and compact ejecta. 
For this reason, the model assumes uniform properties for the matter in the colliding shells and 
treats their evolution self-similarly. In other words, evolution of shocks and their synchrotron 
emission depend only on time or equivalently distance from the center, rather than to both time and 
distance as independent variables. Due to these simplifications, the formalism cannot take into 
account lateral variation of properties in the jet/outflow. However, as suggested in the 
literature~\cite{gw170817cocoon,gw170817cocoon0,gw170817lateradio,gw170817latefasttail,gw170817latexraydecline} 
and we give more arguments in its favour in this review, the observed late afterglows of 
GW/GRB~170817A might have been produced by a continuous flow with a time varying profile of density 
and Lorentz factor. To take into account these variations we simulate afterglows as being generated 
by a composite jet with different density and speed for each component. They are simulated separately 
and can be considered as presenting part of the outflow with corresponding properties. We emphasize 
that these components are {\it effective} presentation of sections of a jet or outflow, which due to the 
simplicity of the model could not be properly simulated as a single entity. Therefore, it is meaningless 
to e.g. consider their interaction with each others.

\subsection {Prompt emission} \label{sec:prompt}
The large number of parameters of the model used here does not allow a systematic exploration of the parameter 
space. To facilitate the search for plausible models for GW/GRB~170817A we use simulations of typical 
short GRBs in~\cite{hourigrbmag} as a departure point and adjust parameters, notably physically 
most important ones, namely: $r_0$, $p$, $\gamma_{cut}$, $ \gamma'_0$, $\Gamma$, $Y_e\epsilon_e$, 
$\epsilon_B$, $N'$, $n'_c$ and $|B|$ around their prototype values to obtain an acceptable fit to the 
Fermi-GBM prompt gamma-ray light curve and spectrum. We emphasize that as the exploration of parameter 
space of the model is not systematic, values of parameters for best models should be considered as 
order of magnitude estimation. 

Giving the unusual characteristics of GRB~170817A, we try 3 range of Lorentz factor to see which one 
lead to a satisfactory model. Additionally, this exercise allows to explore the well known parameter 
degeneracy of shock/synchrotron model of gamma-ray bursts. Categories are:
\begin{description}
\item {\bf Ultra-relativistic jet} with $\Gamma \sim \mathcal{O}(100)$~\cite{grblorentz,grblorentz0,grblorentz1}, see also simulations in~\cite{hourigrbmag}; 
\item {\bf Structured relativistic jet} with $\Gamma \sim \mathcal{O}(10)$~\cite{gw170817xray,gw170817rprocess};
\item {\bf Mildly relativistic cocoon} with $\Gamma \sim \mathcal{O}(1)$~\cite{grbcocoon,nakarpoynting,gw170817cocoon,gw170817cocoon0,grboffaxprobab}. 
\end{description}
Here a {\it cocoon} means a mildly relativistic, mildly collimated outflow with a Lorentz factor 
of $\sim 2-3$. Nonetheless, its angular extension far from the core of the ejecta may have smaller Lorentz 
factor. This slow part of the outflow is not relevant for the high energy prompt emission but as we 
show in the next subsection it can be important for the radio afterglow. 

Our tests show that a cocoon model cannot reproduce Fermi data. Therefore, we do not discuss such 
models further in this section. Examples of such models are discussed in~\cite{hourigw170817}. We 
should also remind that these cocoon models are not exactly the same as {\it cocoon break out} 
model~\cite{grbcocoon} suggested by~\cite{gw170817cocoon0,gw170817cocoonsimul}, which is a different 
mechanism, i.e. not exactly a synchrotron emission. In any case, as we show in Sec. \ref{sec:ag} a cocoon 
or cocoon break out or a chocked jet in the cocoon break out alone cannot explain prompt gamma-ray and 
even afterglows, because X-ray afterglow needs a jet with an ultra-relativistic component.

\begin{figure}
\begin{center}
\begin{tabular}{p{14cm}}
\includegraphics[width=12cm]{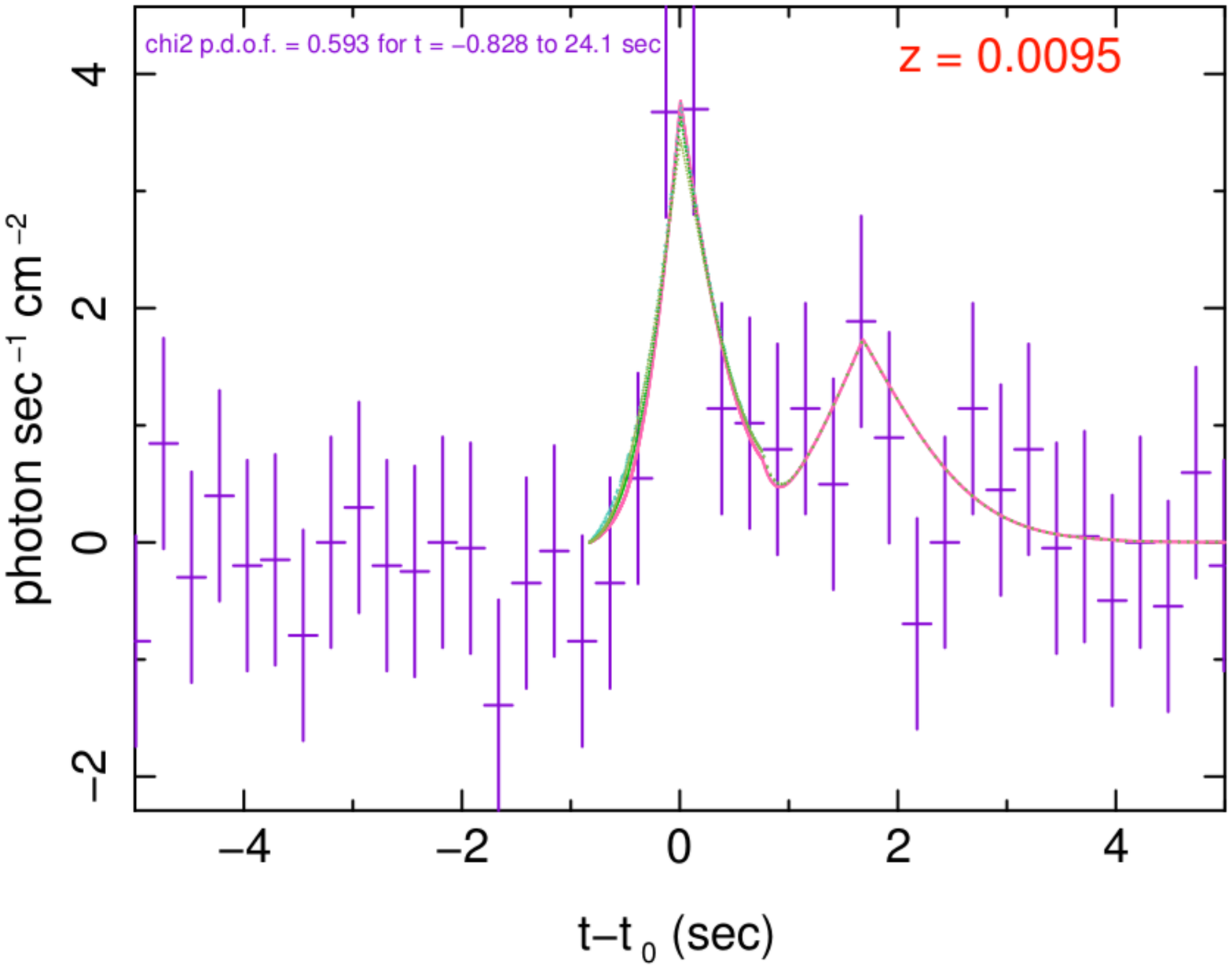} \hspace{-3cm}\includegraphics[width=4cm]{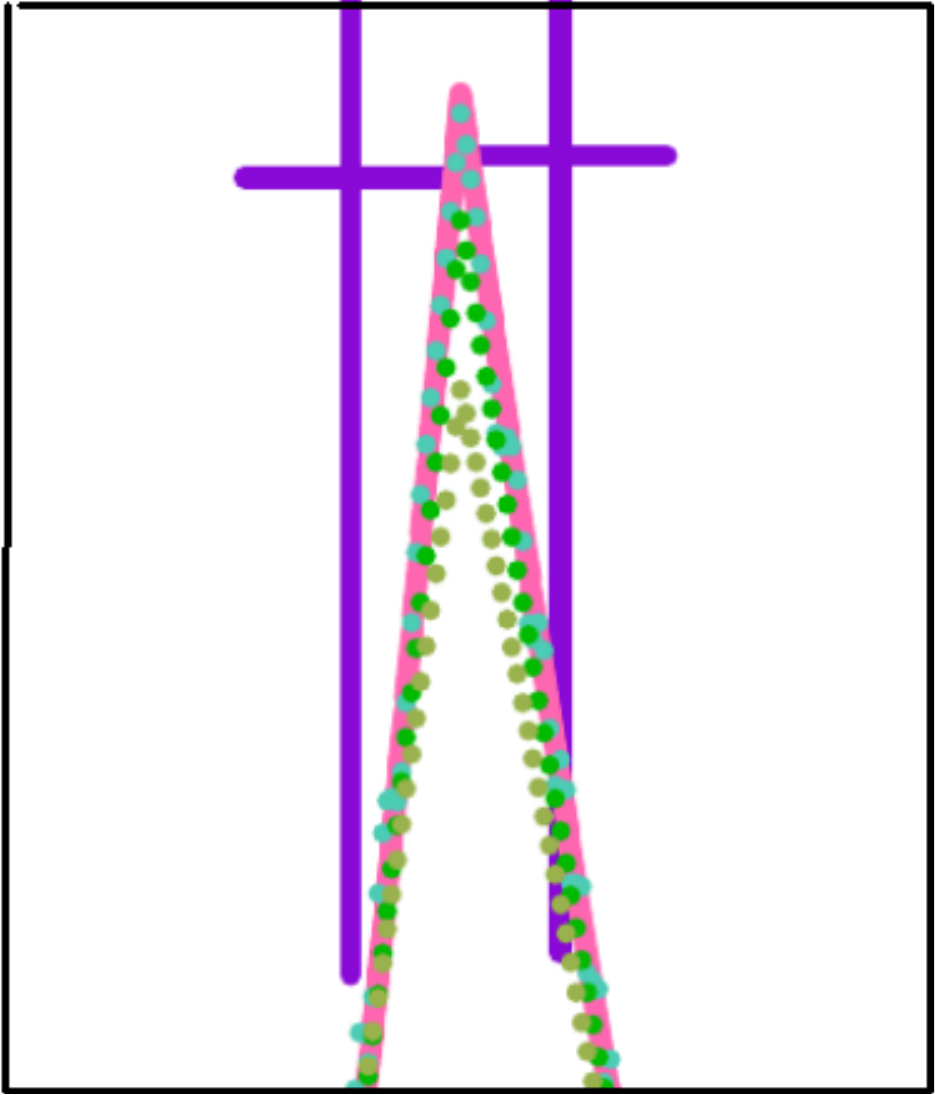}
\end{tabular}
\end{center}
\caption{Light curves of 4 best simulations in 10 keV - 1 MeV. The data is from observations 
of Fermi-GBM~\cite{gw170817fermi}. This plot shows that these simulations have very similar light 
curves. The inset is a zoom on the first peak and shows the slight difference of the amplitude of 
the first peak in these models. The value of $\chi^2$ is for the full line corresponding to 
model No. 2 in Table \ref{tab:param} for the first peak and model No. 3 for the second peak. 
Other curves (doted lines) correspond to model No. 1 with and without an external magnetic field 
(blue and dark green curves, respectively), and an off-axis model with all parameters the same as 
model No. 2, except column density of ejecta which is $n'_c = 5 \times 10^{25}$ cm$^{-2}$ 
(light green). The value of $\chi^2$ per degree of freedom of the first two simulations are about 
$0.02$ larger than model No. 2 and that of the last model is $\sim 0.03$ larger. \label{fig:totlc}}
\end{figure}

\begin{figure}
\begin{center}
\begin{tabular}{p{4.5cm}p{4.5cm}p{4.5cm}}
\hspace{-1cm} \includegraphics[width=8cm]{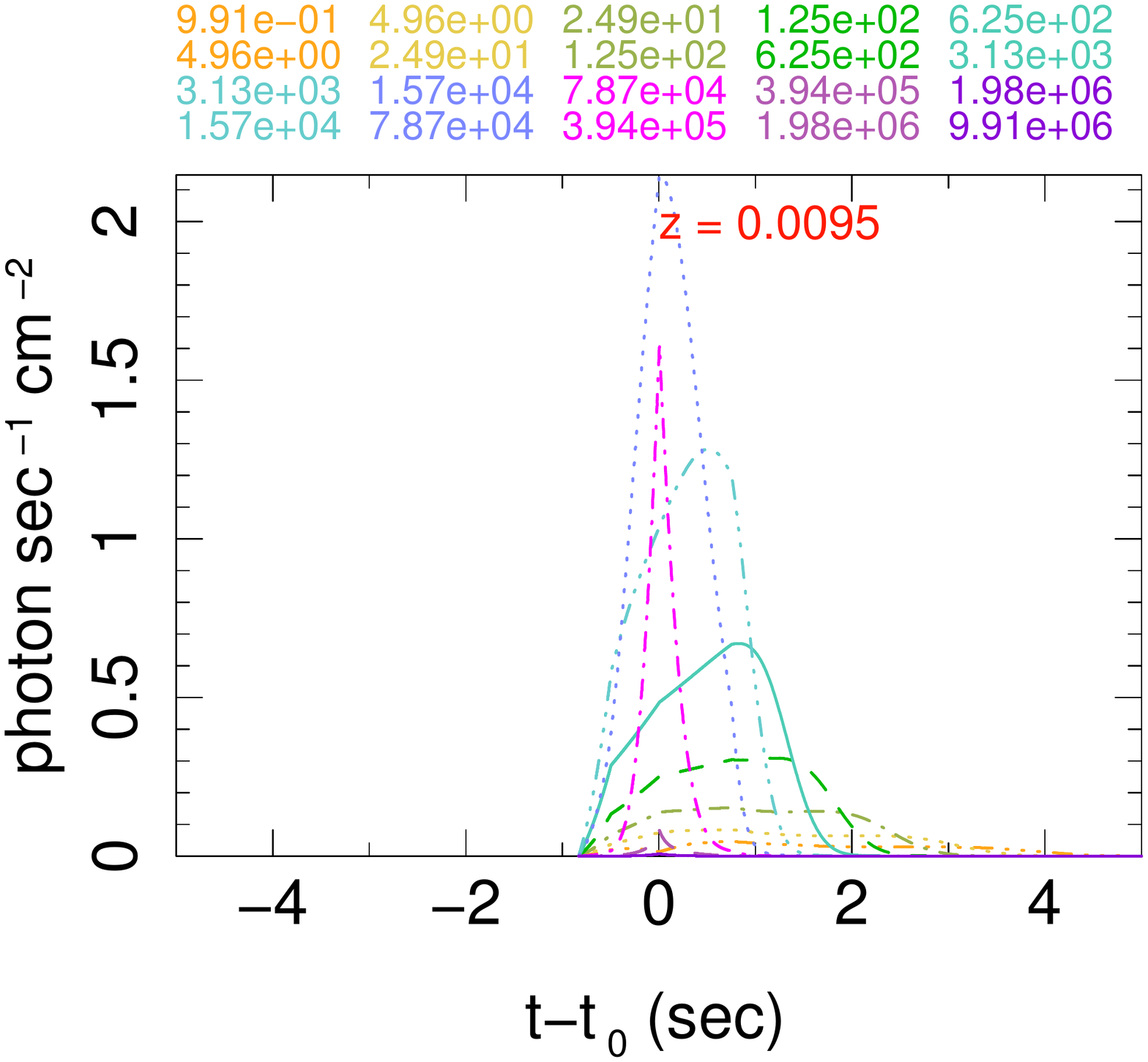} & 
\hspace{-2.5cm} \includegraphics[width=8cm]{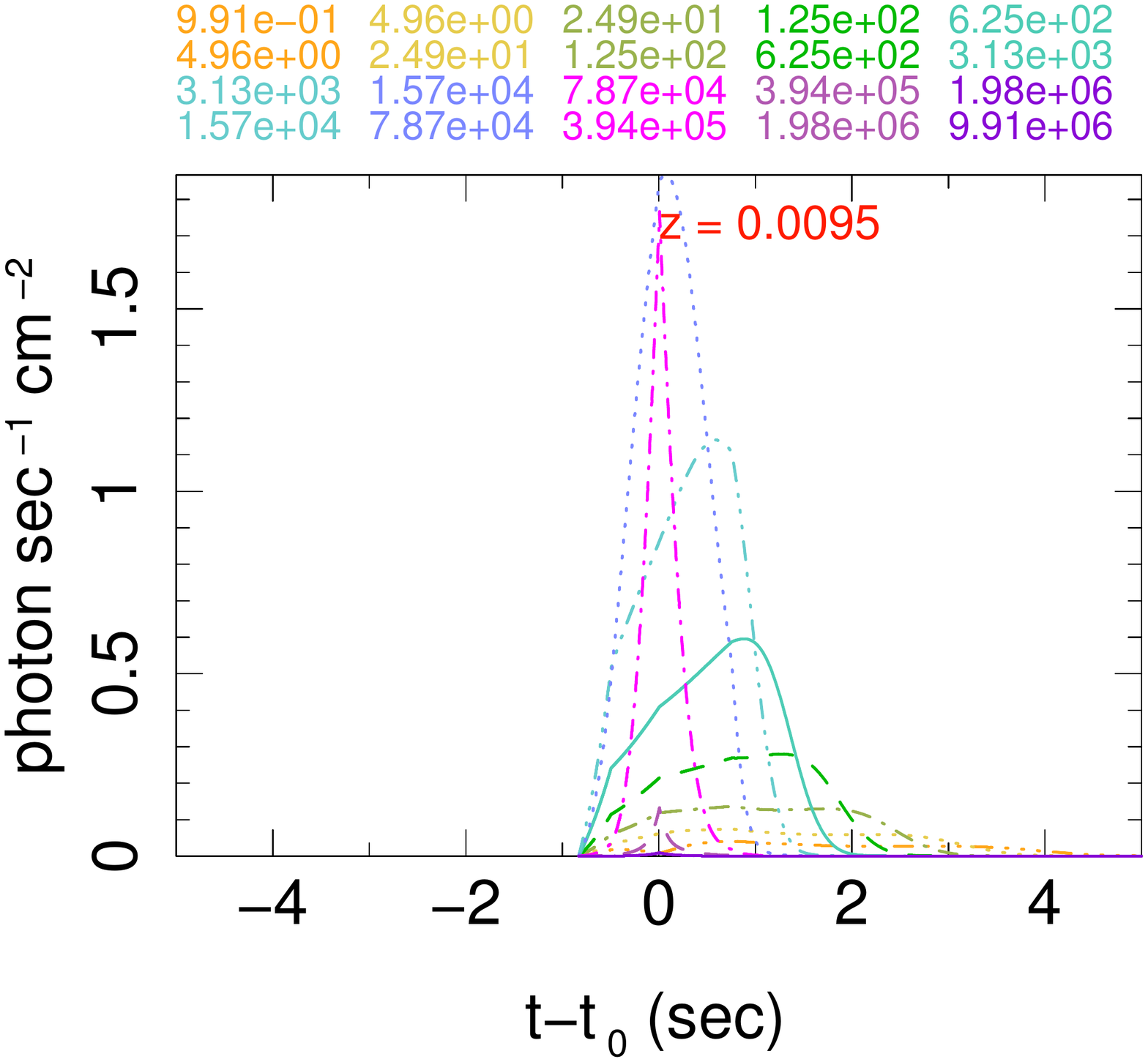} &
\hspace{-2.5cm} \includegraphics[width=8cm]{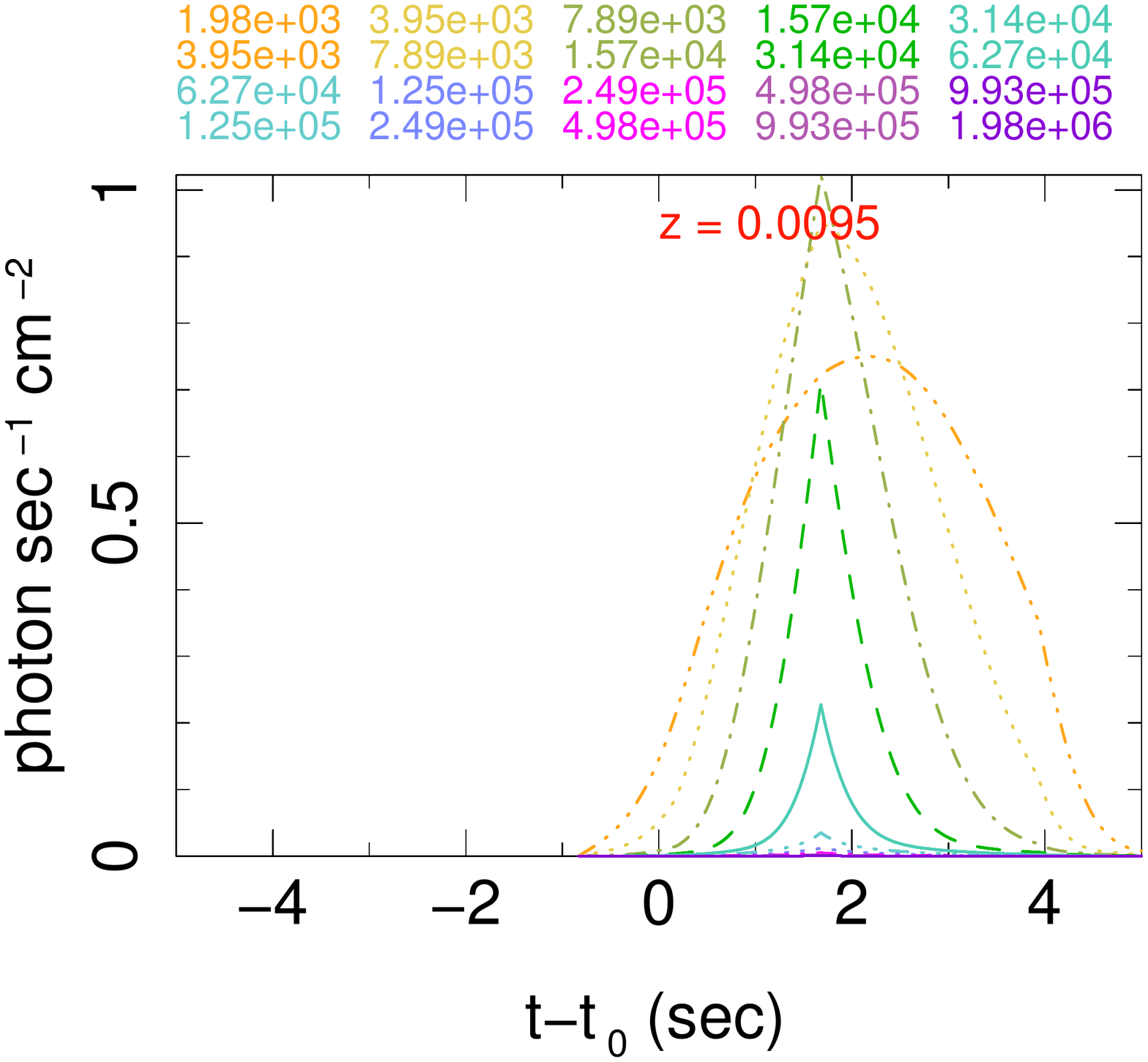}
\end{tabular}
\end{center}
\caption{Light curves of simulated models in energy bands covered by Fermi-GBM and Integral SPI-ACS 
instruments: Left) Simulation No. 2; Center) Simulation No. 1 without external magnetic field; 
Right) Second peak, that is simulation No. 3. All simulation numbers refer to Table \ref{tab:param}. 
Minimum and maximum of each energy band in eV is written in the corresponding color on the top 
of each plot. Notice that the second peak is simulated in lower energy bands than the first 
peak. The lag between highest energy bands is roughly zero and consistent with observation of 
short GRBs.\label{fig:lcbands}}
\end{figure}

Fig. \ref{fig:totlc} shows light curves of 4 simulated bursts with best chi-square fits to 
the Fermi-GBM's 10 keV-1 MeV band data. The two peaks in the observed light curve are simulated 
separately and adjusted in time such that the sum of two peaks minimize the chi-square fit. 
Fig.\ref{fig:lcbands} shows light curves in narrower bands for each peak. Table \ref{tab:param} 
shows the value of parameters for these simulations. Examples of light curves and spectra of 
simulations which were explored to find best fits to the data are given in~\cite{hourigw170817} and 
interested readers can refer to that work. In addition, those simulations were necessary to understand 
degeneracy of the model and how they might impact interpretation of the results.

\begin{figure}
\begin{center}
\begin{tabular}{p{6cm}p{6cm}}
a) \includegraphics[width=8cm]{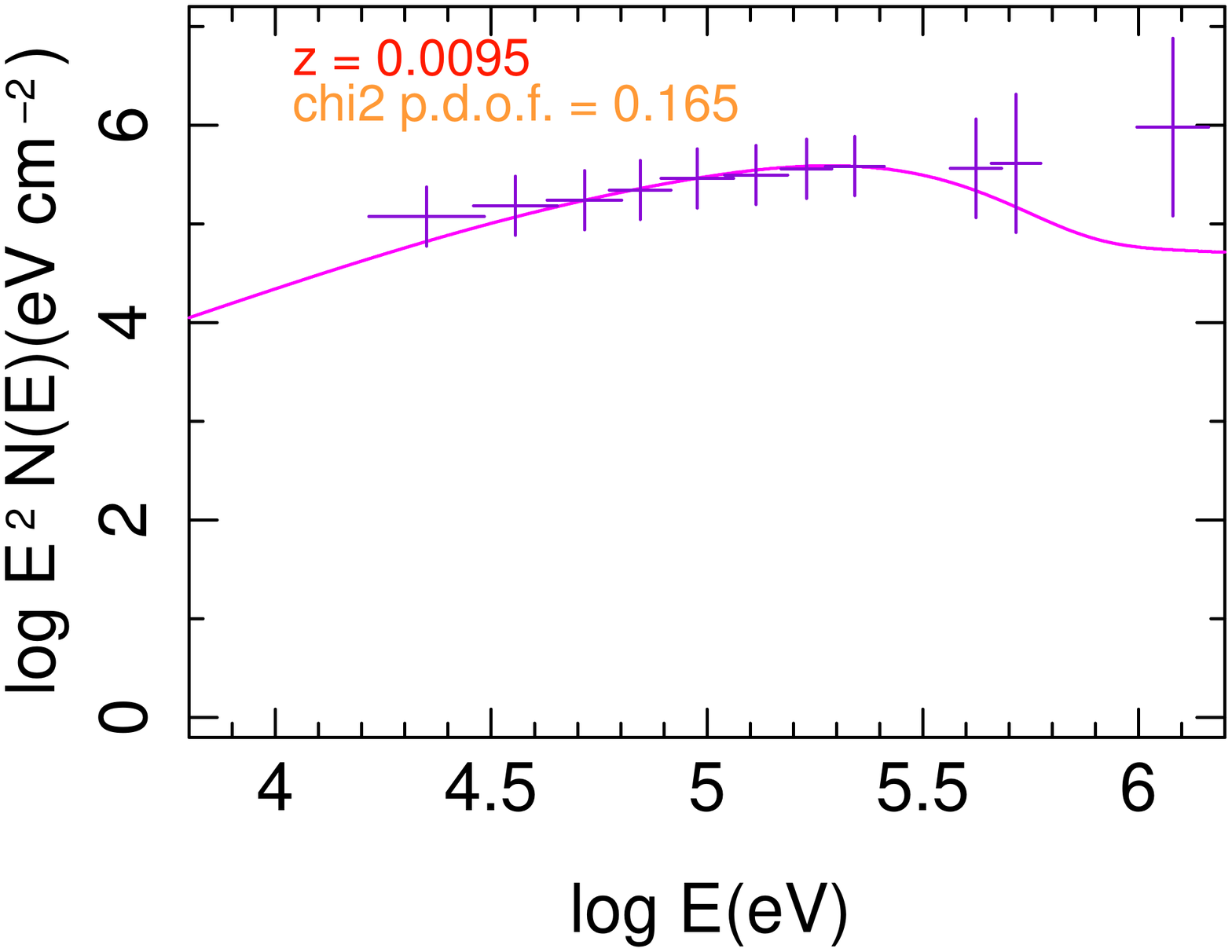} & 
b) \includegraphics[width=8cm]{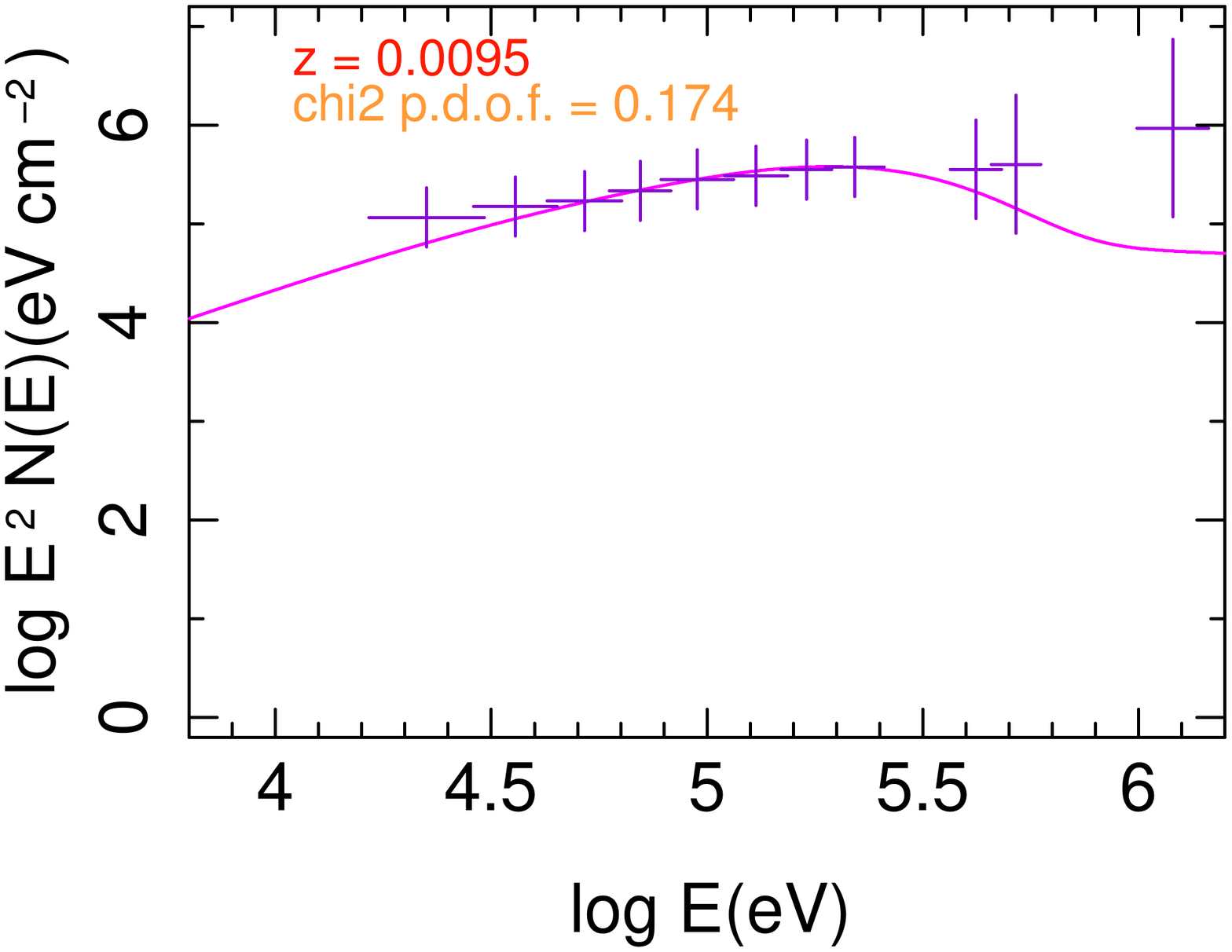}  \\
c) \includegraphics[width=8cm]{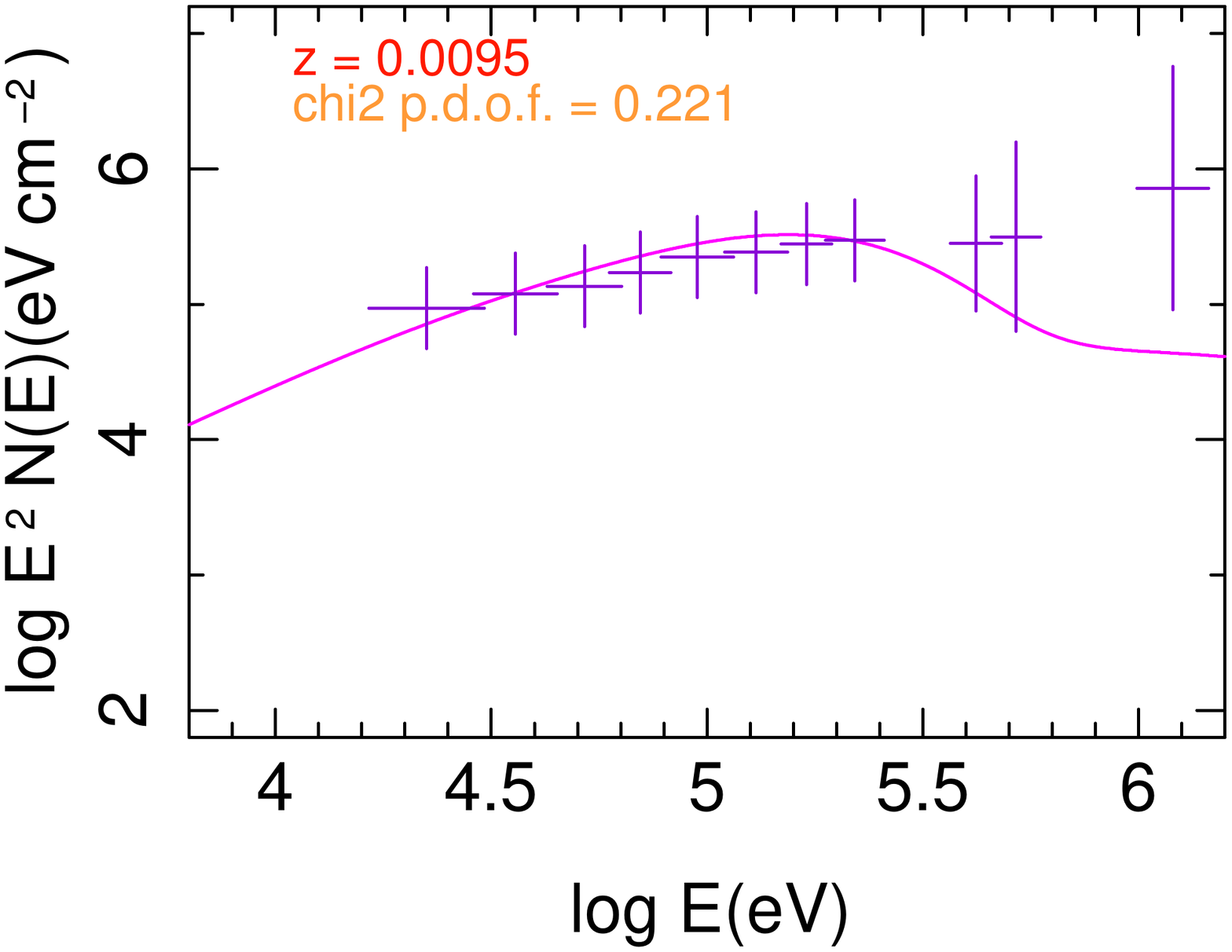} &
d) \includegraphics[width=8cm]{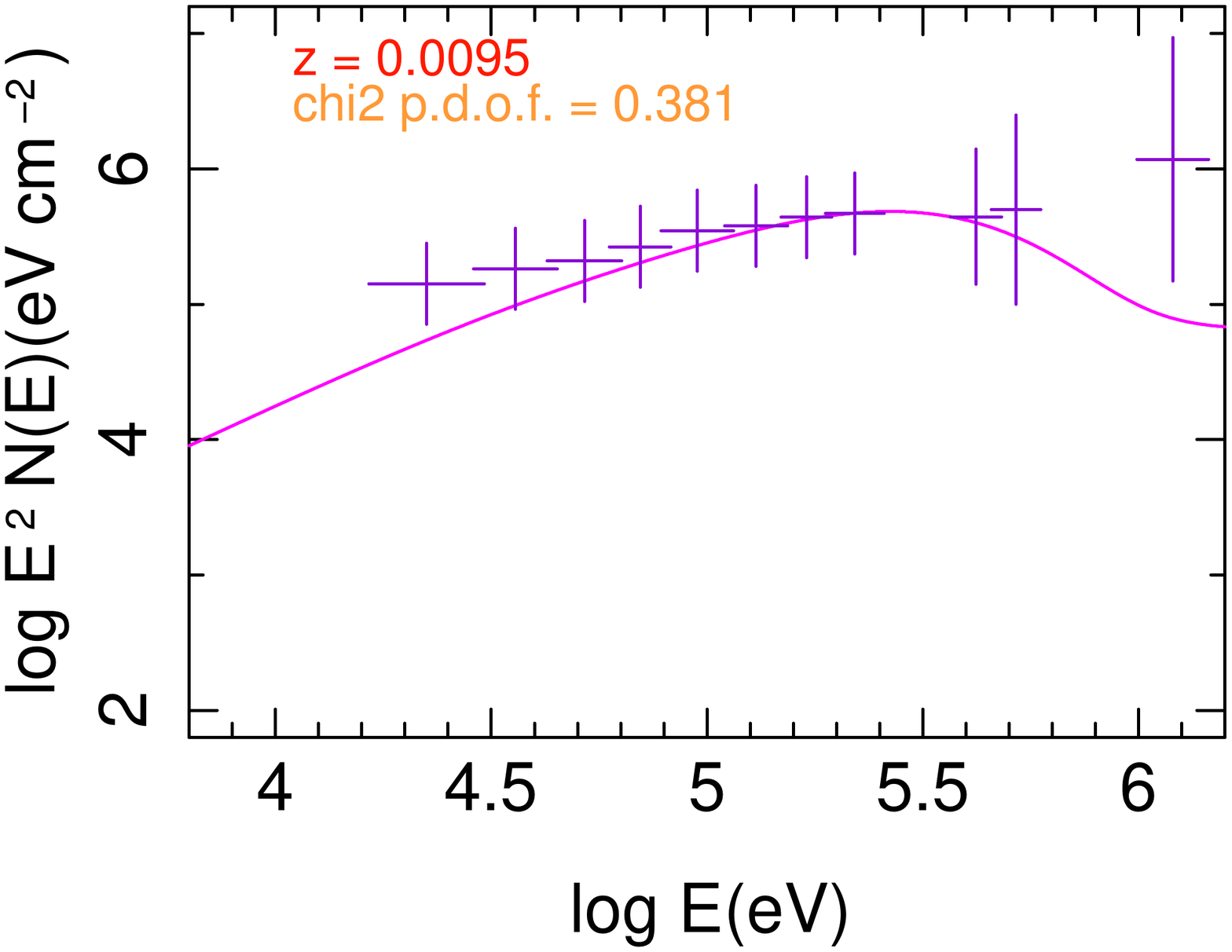} 
\end{tabular}
\end{center}
\caption{Spectra of simulated models fitted to Fermi-GBM data: a) Model No. 1; b) Model No. 1 
without external magnetic field; c) Model No. 2; d) A model with the same parameters as models No. 2 
except for $n'_c = 5 \times 10^{25}$ cm$^{-2}$. As the published spectral data in~\cite{gw170817fermi} 
is in count rate, after changing it to energy flux we used peak energy from~\cite{gw170817fermi} to 
normalize the data such that at $E = E_{peak} = 215 \pm 54$~keV observed and simulated spectra have the 
same amplitude. For this reason, spectra of simulated models have much smaller $\chi^2$ than their 
corresponding light curves. \label{fig:spect}}
\end{figure}

Degeneracy of light curve can be partially resolved by fitting the spectrum of the first peak to 
data, shown in Fig. \ref{fig:spect}. We do not fit spectrum of the second peak because Fermi 
spectrum of this peak~\cite{gw170817fermi} includes only 2 measured points at lowest 
energies and the rest are upper limits. It is evident that model d) in Fig. \ref{fig:spect} has 
a weaker fit to data than others. However, in view of uncertainties of the data, their 
differences are too small to provide a statistically significant criteria in favour of one of these 
models. Nonetheless, comparison of spectra in \ref{fig:spect}-a and \ref{fig:spect}-b, which 
their only difference is an external magnetic field in the former, may be interpreted as the 
necessity of a weak magnetic field in addition to the field induced by Fermi processes in 
the shock front. Indeed, in the next section we show that the remnant of the jet after internal 
shocks had preserved its collimation and coherence well after prompt shocks. Therefore, 
it had to have intrinsic magnetic field. As for degeneracies, they can be ultimately removed 
when the same formalism is used to analyse afterglows of this burst.

\newcounter{x}
\setcounter{x}{\thetable}
\begin{table}
\begin{center}
\caption{Parameter set of simulated prompt emission models \label{tab:param}}
\end{center}
{\scriptsize
\begin{center}
\begin{tabular}{p{2.5cm}p{10mm}p{10mm}p{12mm}p{12mm}p{10mm}p{12mm}p{10mm}p{12mm}p{10mm}p{10mm}p{10mm}}
\hline
Model Descr. & ~mod. & $\Gamma$ & $r_0$ (cm) & $\frac{\Delta r_0}{r_0}$ & $(\frac{r}{r_0})_{max}$ & $p$ & $\gamma_{cut}$ & $\kappa$ & $ \gamma'_0$ & $ \tau$ & $ \delta$  \\
\hline 
\multirow{4}{2.5cm}{~1:~GW/GRB~170817: first peak, rel.jet} & ~1 & 100 & $2\times10^{10}$ & $5\times10^{-5}$ & 1.5 & 2.5 & 10 & 0 & 1.5 & - & 1 \\
 & ~0 & - & - & - & 1.5 & - & 10 & 0 & - & 0 & - \\
 & ~2 & - & - & - & 1.5 & - & 10 & 0 & - & - & 3 \\
 & ~2 & - & - & - & 4 & - & 10 & 0 & - & - & 5 \\
\hline
\multirow{4}{2.5cm}{~2:~GW/GRB~170817: first peak, off-axis} & ~1 & 10 & $2\times10^{10}$ & $5\times10^{-5}$ & 1.5 & 2.5 & 10 & 0 & 1.5 & - & 1 \\
 & ~0 & - & - & - & 1.5 & - & 10 & 0 & - & 0 & - \\
 & ~2 & - & - & - & 1.5 & - & 10 & 0 & - & - & 3 \\
 & ~2 & - & - & - & 4 & - & 10 & 0 & - & - & 5 \\      
\hline
\multirow{4}{2.5cm}{~3:~GW/GRB~170817: second peak} & ~1 & 30 & $6\times10^{10}$ & $5\times10^{-5}$ & 1.5 & 2.5 & 10 & 0 & 1.5 & - & 1 \\
 & ~0 & - & - & - & 1.5 & - & 10 & 0 & - & 0 & - \\
 & ~2 & - & - & - & 1.5 & - & 10 & 0 & - & - & 3 \\
 & ~2 & - & - & - & 4 & - & 10 & 0 & - & - & 5 \\
\hline
\end{tabular}
\end{center}
}

\setcounter{table}{\thex}
\begin{center}
\caption{{\bf(continued)} Parameter set of simulated prompt emission models}
\end{center}
{\scriptsize
\begin{center}
\begin{tabular}{p{2.5cm}p{12mm}p{10mm}p{10mm}p{10mm}p{20mm}p{20mm}p{15mm}p{12mm}p{12mm}p{12mm}}
\hline
Model Descr. & $\epsilon_B$ & $\alpha_B$ & $\epsilon_e Y_e$ & $\alpha_e$ & $N'$ (cm$^{-3}$) & $n'_c$ (cm$^{-2}$) & $|B|$ (kG) & $f$ (Hz) & $\alpha_x$ & $\phi$(rad.)\\
\hline
\multirow{4}{2.5cm}{~1:~GW/GRB~170817: first peak, rel.jet} & $10^{-4}$ & -1 & 0.01 & -1 & $2 \times 10^{14}$ & $10^{25}$ & 0.8 & 500 & - & - \\
 & - & -2 & - & -2 & - & - & - & - & 1 & - \\
 & - &  2 & - &  2 & - & - & - & - & 2 & - \\
 & - &  4 & - &  4 & - & - & - & - & 3 & - \\
\hline
\multirow{4}{2.5cm}{~2:~GW/GRB~170817: first peak, off-axis} & $10^{-4}$ & -1 & 0.03 & -1 & $2 \times 10^{14}$ & $5 \times 10^{24}$ & 0.5 & 500 & 1 & - \\
 & - & -2 & - & -2 & - & - & - & - & 1 & - \\
 & - &  2 & - &  2 & - & - & - & - & 2 & - \\
 & - &  4 & - &  4 & - & - & - & - & 3 & - \\
\hline
\multirow{4}{2.5cm}{~3:~GW/GRB~170817: second peak} & $10^{-4}$ & -1 & 0.01 & -1 & $2 \times 10^{13}$ & $5 \times 10^{23}$ & 0 & - & - & - \\
 & - & -2 & - & -2 & - & - & - & - & - & - \\
 & - &  2 & - &  2 & - & - & - & - & - & - \\
 & - &  4 & - &  4 & - & - & - & - & - & - \\
\hline
\end{tabular}
\end{center}
}
\begin{description}
\item {$\star$} Each data line corresponds to one simulated regime, during which quantities listed 
here remain constant or evolve dynamically according to fixed rules. A full simulation of a burst 
usually includes multiple regimes (at least two). 
\item {$\star$} Horizontal black lines separate time intervals (regimes) of independent simulations 
identified by the number shown in the first column.
\item {$\star$} A dash as value for a parameter presents one of the following cases: it is 
irrelevant for the model; it is evolved from its initial value according to an evolution equations 
described in~\cite{hourigrb,hourigrbmag}; or it is kept constant during all regimes. 
\end{description}
\end{table}
Parameters of the best models of the prompt emission show that in comparison with other short GRBs 
with accompanying kilonova such as GRB~130603B~\cite{grb130603bkilonova,grb130603bkilonova0} Lorentz 
factor and densities of colliding shells of GRB~170817 in our direction was at least a few folds 
smaller, see~\cite{hourigw170817} for details. Although off-axis view of the jet, as we will discuss 
further in the next subsection, may be somehow responsible for the weakness of this burst, there is 
evidence for the involvement of intrinsic characteristics. This subject will be more detailed in 
Sec. \ref{sec:progenitor}.

\subsection {Late afterglows} \label{sec:ag}
As we discussed in Sec. \ref{sec:obs} electromagnetic afterglows of GW/GRB~170817A were the most 
unusual among both short and long GRBs. Notably, long lasting brightening had never been observed in 
any other GRB. It is however useful to remind that at present no other GRB had follow up observations 
for as long as this burst. In particular, follow up of other short GRBs have been limited to just 
a few days. Thus, we ignore how exceptional is the behaviour of afterglows of GW/GRB~170817A. 

Although, off-axis view of a structured jet predicts late brightening of 
afterglows~\cite{grboffaxis,gw170817latexraystructjet,gw170817latexary,grboffaxprobab}, the observed 
decline of X-ray flux at $\lesssim T+134$ days~\cite{gw170817latexraydecline} is 
inconsistent with simulations of significantly off-axis emission~\cite{gw170817latexraystructjet}, 
which predict a break after a few hundred days. Other simulations, for instance those reported 
by~\cite{gw170817latexary,gw170817offaxecocoondisc} predict earlier break, but cannot discriminate 
between off-axis structured jet and cocoon models~\cite{gw170817lateradio} and need polarimetry 
and imaging to discriminate between them~\cite{gw170817offaxecocoondisc}. In any case, initially it 
seemed reasonable to assume that at late times the section of the jet along our line of sight, which even 
before prompt shocks was not as dense and boosted as in typical short GRBs, had to have dissipated 
its energy, and its Lorentz factor and density should have been decreased to negligibly small values. 
In this case the late brightening had to have another origin, for instance MHD instabilities leading 
to an increase in magnetic energy dissipation~\cite{grbjetsimulinstab,grbjetsimulinstab0}; external 
shock on the ISM/circumburst material of a mildly relativistic thermal cocoon ejected at the same 
time as the dissipated relativistic GRB making jet~\cite{nakarpoynting,cocoon}; late outflows from 
an accretion disk~\cite{latexrayexcess}; or decay of isotopes in the 
kilonova~\cite{hourigw170817ag,kilonovadecay}. Therefore, it was not certain that late afterglows 
were directly related to the relativistic jet models concluded from analysis of the prompt emission, 
see also the commentary of~\cite{latexcomment} on the issue of finding a reliable explanation for 
this exceptional discovery.

Here we first discuss simulation of external shocks of a mildly relativistic outflow on the 
ISM/circum-burst material in the framework of formalism reviewed in Sec. \ref{sec:model} and highlight 
shortcomings of such models for afterglows of GW/GRB 170817A. Then, we discuss models which can 
explain the data.

\subsubsection{Shortcomings of a mildly relativistic outflow}
\begin{figure}
\begin{center}
\begin{tabular}{p{5cm}p{5cm}p{5cm}p{5cm}}
\hspace{-1cm}\includegraphics[width=6cm]{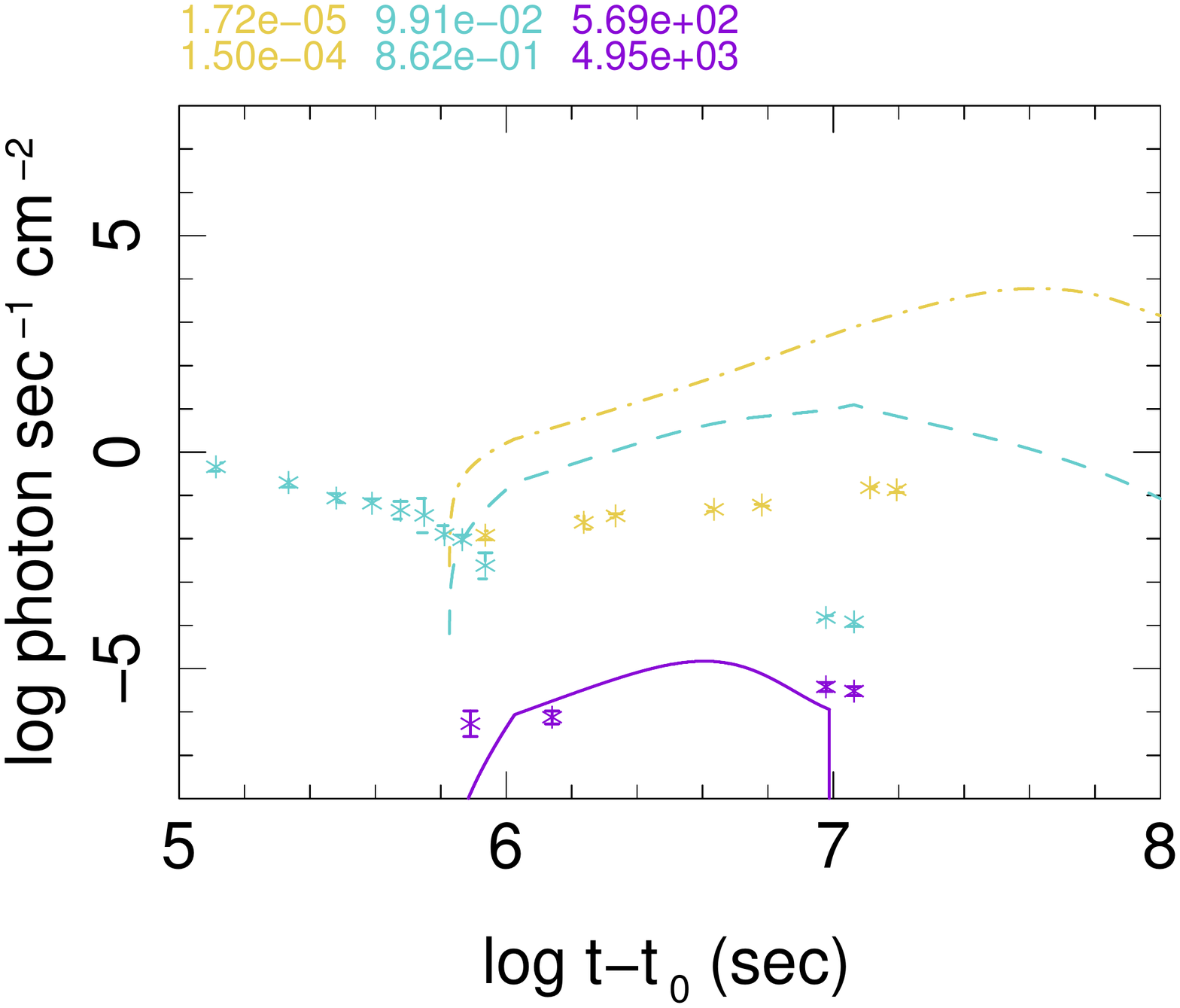} & 
\hspace{-1.8cm}\includegraphics[width=6cm]{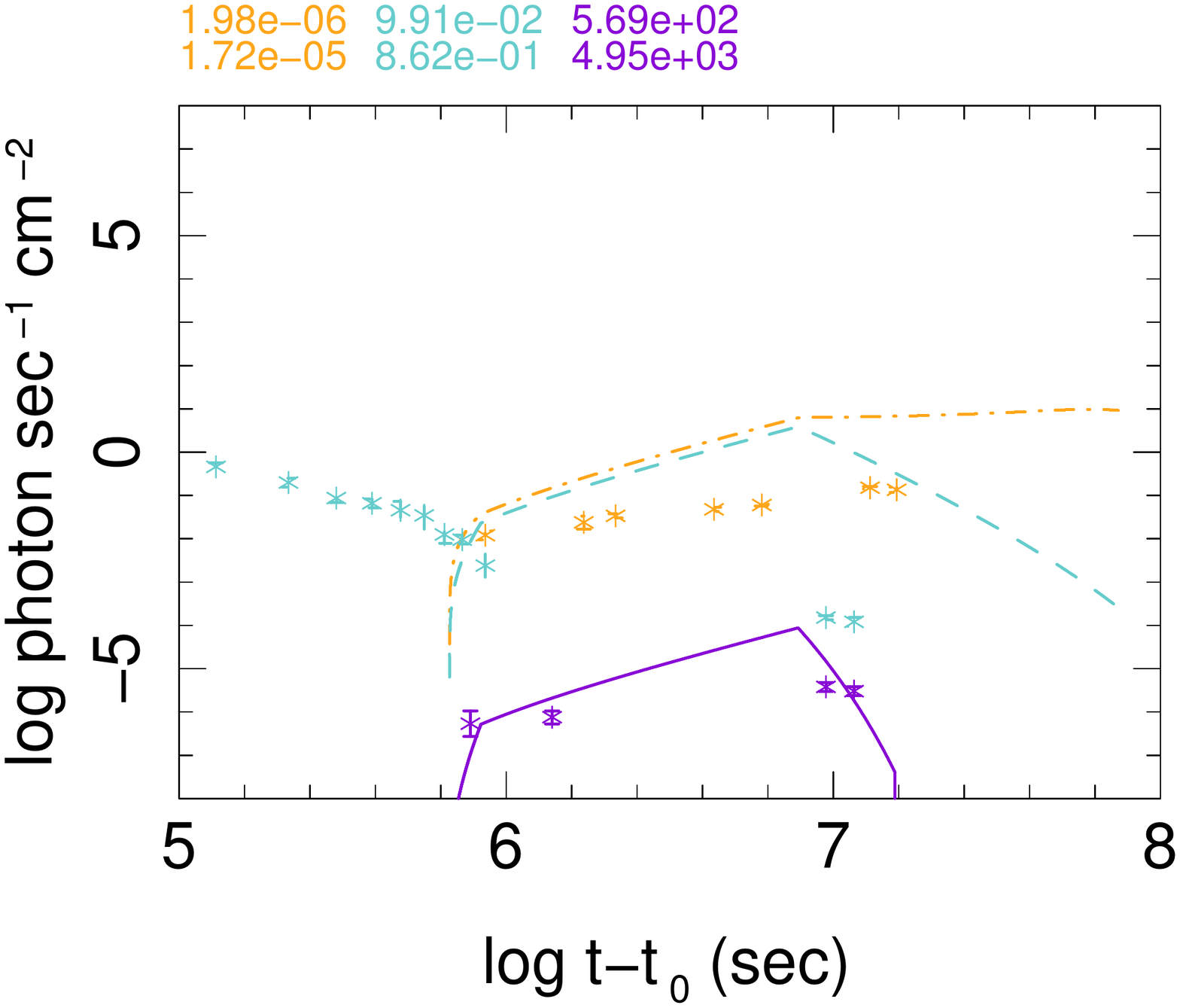} & 
\hspace{-1.8cm}\includegraphics[width=6cm]{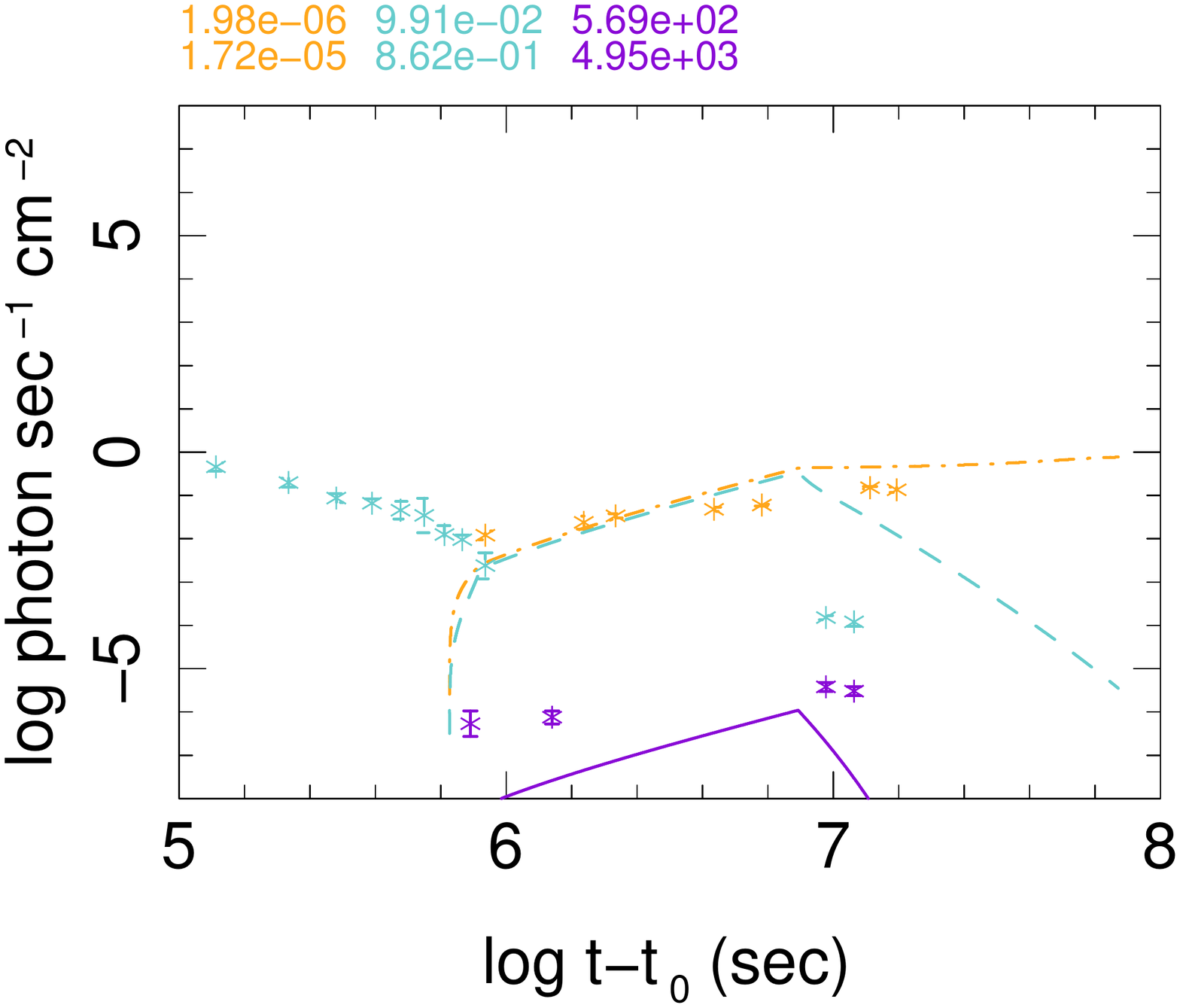} & 
\hspace{-1.8cm}\includegraphics[width=6cm]{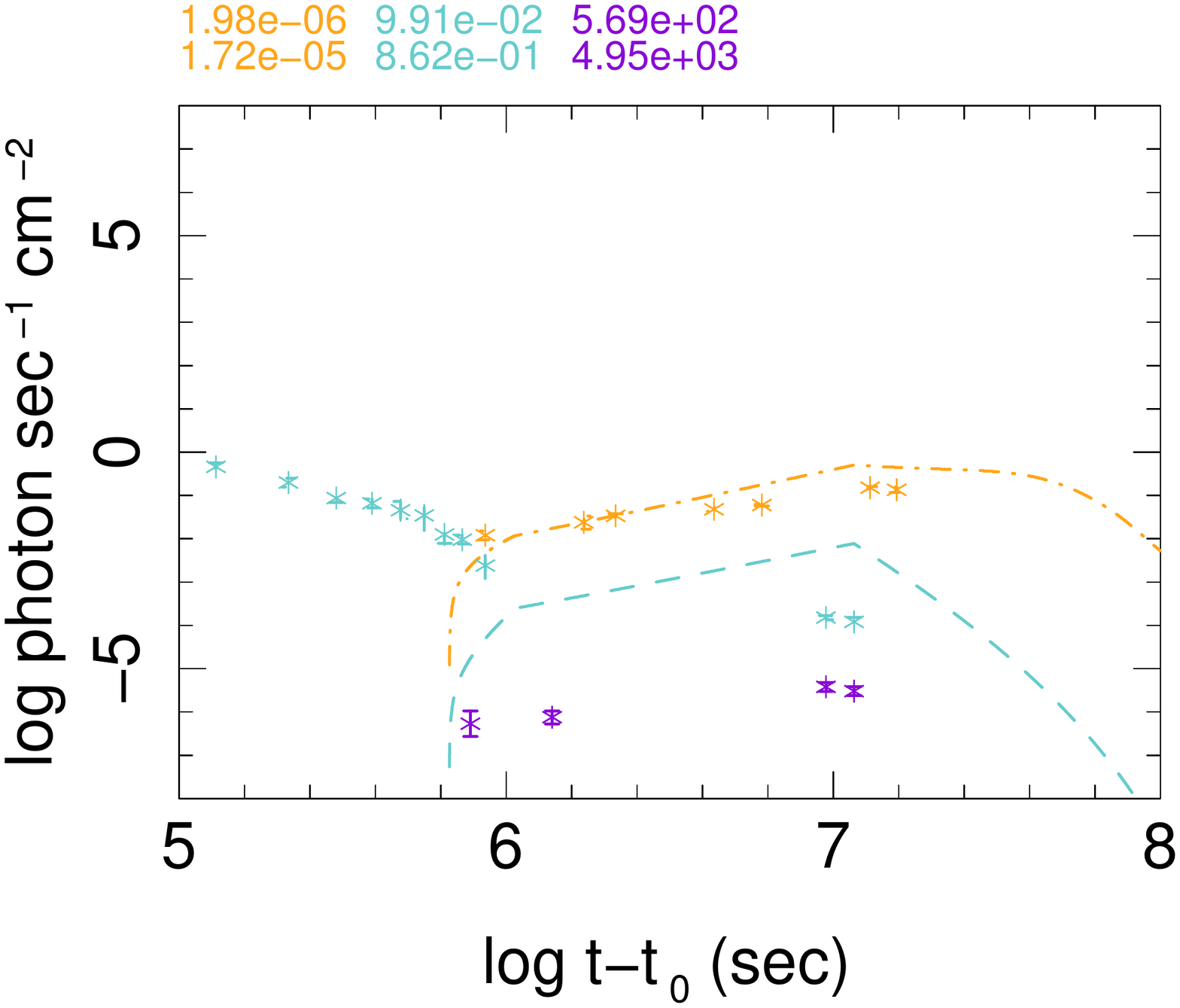} \\
a) & b) & c) & d)
\end{tabular}
\caption{X-ray, optical, and radio light curves of simulated models without taking into account 
synchrotron self-absorption. Stars present data taken 
from:~\cite{gw170817latexary,gw170817xraycxc260} (X-ray), 
\cite{gw170817optdes,gw170817rprocess,gw170817lateopt,gw170817latexraydecline} (optical), 
\cite{gw170817earlyradio1,gw170817earlyradio,lateradio} (radio). Model a) has $\Gamma = 2.3$ and 
others have $\Gamma = 1.2$. Complete list of parameters for these simulations can be found 
in~\cite{hourigw170817ag}. If it is assumed that the centroid of outflow is oblique with respect 
to the line of sight, the effect of projection must be taken into account and the Lorentz factor 
should be considered to be larger by a factor of $1/\cos \theta_c$, where $\theta_c$ is the angle 
between centroid of outflow and the line of sight. \label{fig:lclowgamma}}
\end{center}
\end{figure}

Irrespective of energy dissipation mechanism, most of explanations described in the previous 
paragraphs predict a mildly relativistic outflow with a Lorentz factor $\Gamma < 5$ as the ultimate 
origin of afterglows. Indeed, even one of the best prompt models shown in Figs. \ref{fig:totlc} and 
\ref{fig:spect} has a Lorentz factor $\Gamma \sim \mathcal{O}(10)$. Energy dissipation, e.g. through 
weak internal shocks and interaction with material close to the merger could easily reduce jet's Lorentz 
factor to $\Gamma \sim 2-3$, as had been suggested in the early literature on afterglows of GW/GRB~170817A. 
Fig. \ref{fig:lclowgamma} shows a few examples of synchrotron emission from external shocks of such 
outflows. It is clear that none of them can fit all the multi-band data. Although models in 
Fig. \ref{fig:lclowgamma}-a \& b fit well X-ray data, they over-produce both optical and radio 
emission. By contrast, models c) and d) in this figure have an acceptable fit to radio data, but not 
enough X-ray and both of them over-produce optical emission by a large amount. 

Regarding performance of these simulation, several issues need more clarification. It is well known 
that synchrotron self-absorption of radio emission can be important. In this case, one can assume 
that most of radio emission in the model a) which has higher Lorentz factor is absorbed and only 
radio from a slower and less dense section of outflow, presented by models c) and d), can escape 
the shocked region. Examples of the effect 
of self-absorption are presented in~\cite{hourigw170817ag}. On the other hand, this explanation 
cannot solve the problem of over-production of optical photons. Although the simulated optical band 
in Fig. \ref{fig:lclowgamma} is wider than HST F606W and SDSS r' filters or equivalents used for 
acquiring data shown in these plots\footnote{For the time being the number of bands in the simulation 
code is fixed. Therefore, to cover a broad range of energies, from radio to X-ray, width of 
individual bands should be large.}, it is unlikely that multiple orders of magnitude deviation 
between these models and data can be due to the wider simulated band. Moreover, it is crucial to 
remind that optical data is dominated by kilonova mission, which has very different origin and 
spectrum. Therefore, the optical data must be considered as an upper limit to any contribution from 
the GRB. This makes models shown in Fig. \ref{fig:lclowgamma} even more distant from data, unless 
optical photons is strongly observed. Alternatively, one can make simulations consistent with 
optical data, but an additional source of X-ray would be necessary to explain these observations.

The outline of this exercise is that the assumption about extreme energy dissipation in the jet 
after prompt emission is not realistic. Indeed, to increase the contribution of X-ray emission and 
decrease the number of soft photons the spectrum of synchrotron emission must be harder, i.e. its 
peak must be pushed to higher energies such that low energy emissions fall on the fast declining 
side of the spectrum, see examples of synchrotron/self-Compton spectra of GRBs 
in~\cite{hourigrbmag,hourigrbrev}. On the other hand, realization of this requirement needs larger 
Lorentz factors. Moreover, the fact that combination of two simulations 
gives a better fit to data than each component alone reminds us that the formulation of~\cite{hourigrb} 
assumes a jet with uniform Lorentz factor and density. But numerical simulations of ejecta from 
BNS~\cite{nstarbhmergsimul2,nsmergerrprocsimul0,nstarmergeejecta,nstarmergespingw,nstarmergespingw} 
and its further acceleration by transfer of Poynting to kinetic energy~\cite{grbjetsimul} shows that 
both these processes are inhomogeneous. Consequently, the final relativistic jet has a profile 
with varying $\Gamma$, density and extension. Another crucial property of emissions from a 
relativistic jet is their beaming for a far observer. Due to this relativistic effect only an angle 
$\theta_{max} = \arcsin (1/\Gamma)$ around the line of sight is visible to the observer. As we 
discussed in Sec. \ref{sec:prompt} at the time of prompt emission the jet is ultra-relativistic and 
$\Gamma \gtrsim \mathcal {O}(100)$. Therefore, beaming allows to see only emission from a small part 
of the jet along the line of sight. Consequently, anisotropy of the jet is not visible, specially when only 
a relatively narrow gamma-ray band is observed. By contrast, afterglows are usually observed in broad 
bands - from gamma and X-rays to radio - and low energy emissions from side lobes - high latitudes - 
of the jet would be visible and important.

\subsubsection{Multi-component models}
\begin{center}
\begin{figure}
\begin{tabular}{p{9cm}p{9cm}}
{\bf a)} & {\bf b)} \\
\includegraphics[width=9cm]{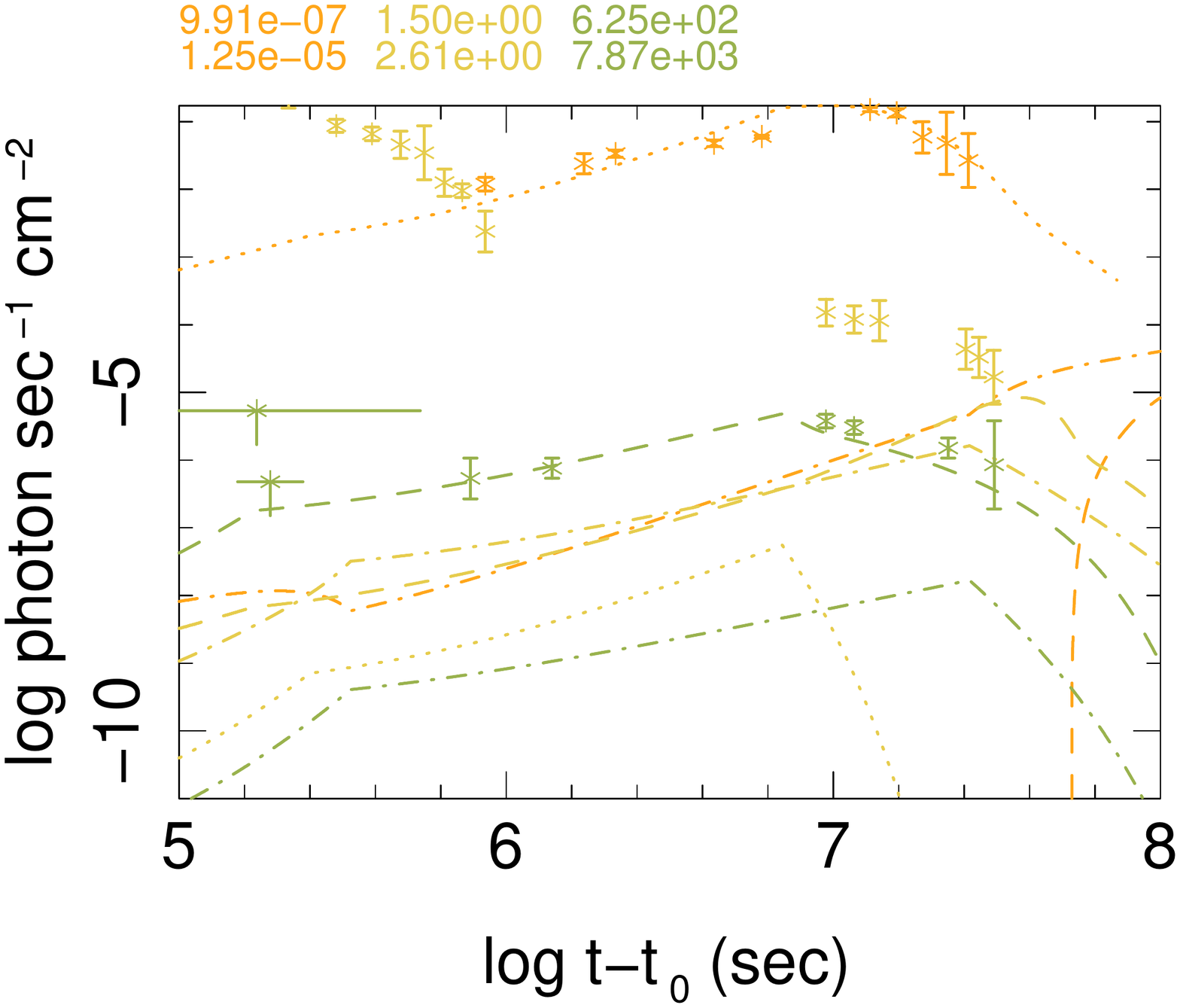} &
\hspace{-1.5cm}\includegraphics[width=9cm]{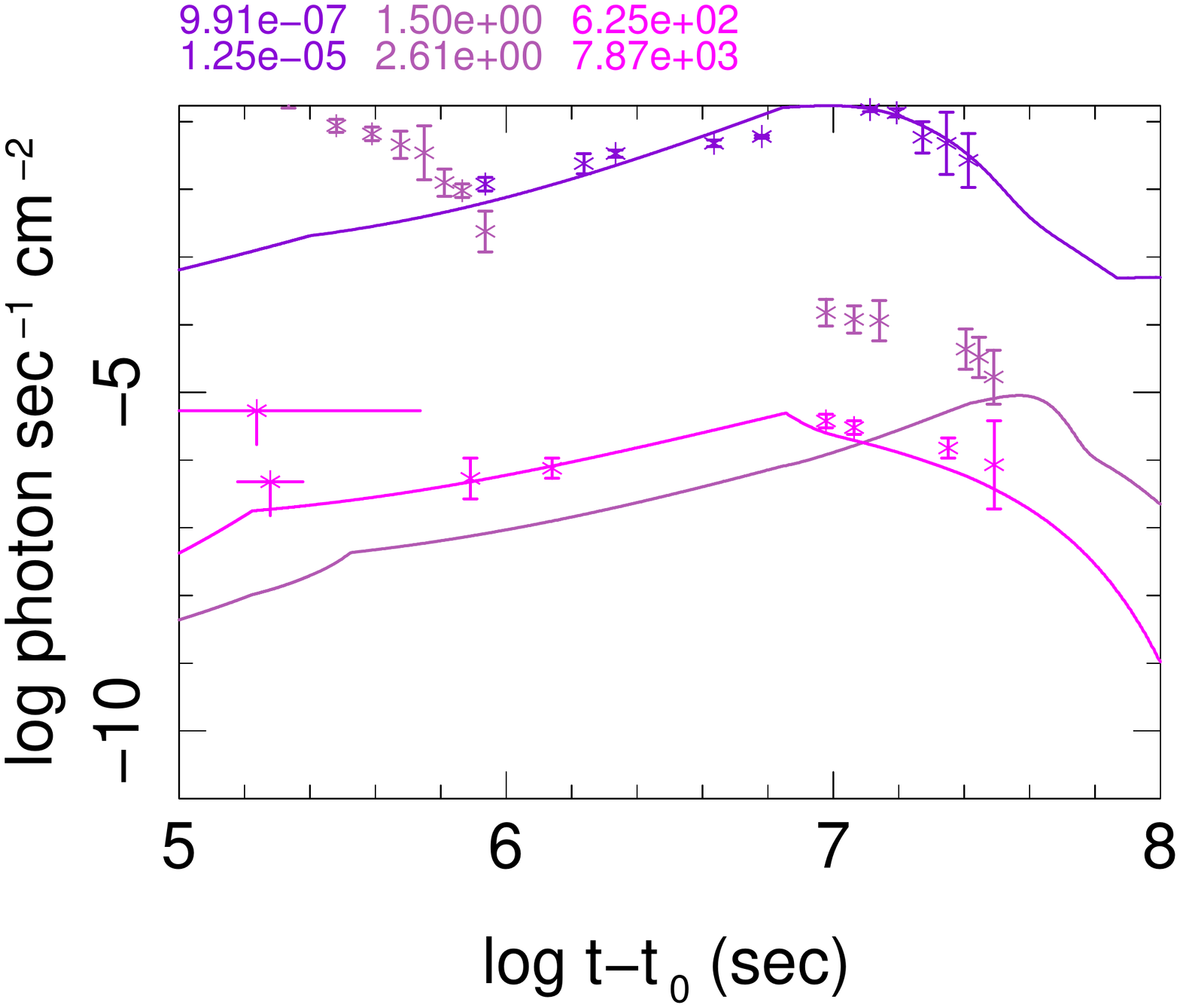}
\end{tabular}
\caption{{\bf a)}: Radio, optical/IR, and X-ray light curves of simulated 3-component. Left: Light 
curves of the 3 components: ultra-relativistic (C1) (dash lines), relativistic (C2) (dash-dot), mildly 
relativistic (C3) (dotted lines). The energy range for each band is written on the top of each plot 
in the same colour/gray scale as the curves. Stars present data taken 
from:~\cite{gw170817latexary,gw170817xraycxc260,gw170817chandraxray260a,gw170817chandraxray358} (X-ray), 
\cite{gw170817optdes,gw170817rprocess,gw170817lateopt,gw170817latexraydecline,gw170817lateopt160,gw170817lateopthstir} (optical), 
\cite{gw170817earlyradio1,gw170817earlyradio,lateradio,gw170817lateradio300} (radio). The upper 
limit of X-ray flux at $<T+10$ is from Neil Gehrels Swift-XRT~\cite{gw170817swiftnustar} and the 
upper limit at $\sim T + 2.2$ is from Chandra observations~\cite{gw170817xray}. 
{\bf b)}: Sum of the light curves of the 3 components: radio (magenta/light grey), optical 
(purple/medium grey), X-ray (dark purple/dark grey). \label{fig:lcagmain}}
\end{figure}
\end{center}

The shortcoming of models simulated in~\cite{hourigrb,hourigrbmag} can be compensated by 
constructing multi-component models in which each component presents approximately an angular 
section of the jet profile with different Lorentz factor, density and extension. In addition, as argued 
in the previous subsection, we allow the presence of an ultra-relativistic component. This strategy, 
leads to a model and a few of its variants which fit X-ray and optical data well and satisfy the upper 
limit constraint imposed by observed optical data~\cite{hourigw170817lateagjet}. Parameters of 
the components of the model are shown in Table \ref{tab:agparam}. Fig. \ref{fig:lcagmain} 
shows light curves of components of this model and their sum in each band. As this model and 
some of its variants shown in Fig. \ref{fig:variants} fit well radio and X-ray data, we presume that their 
predictions for optical emission of the GRB afterglow should be reliable. Under this assumption, these 
models show that after $\sim T+200$~days kilonova emission in optical/IR bands was not anymore 
significant and the afterglow was dominated by synchrotron emission from external shocks of the 
relativistic polar outflow from the merger. 

\begin{center}
\begin{figure}
\begin{tabular}{p{8cm}p{7cm}p{5cm}}
{\bf a)} & \hspace{-1.5cm} {\bf b)} & \hspace{-3.5cm} {\bf c)}\\
\includegraphics[width=8cm]{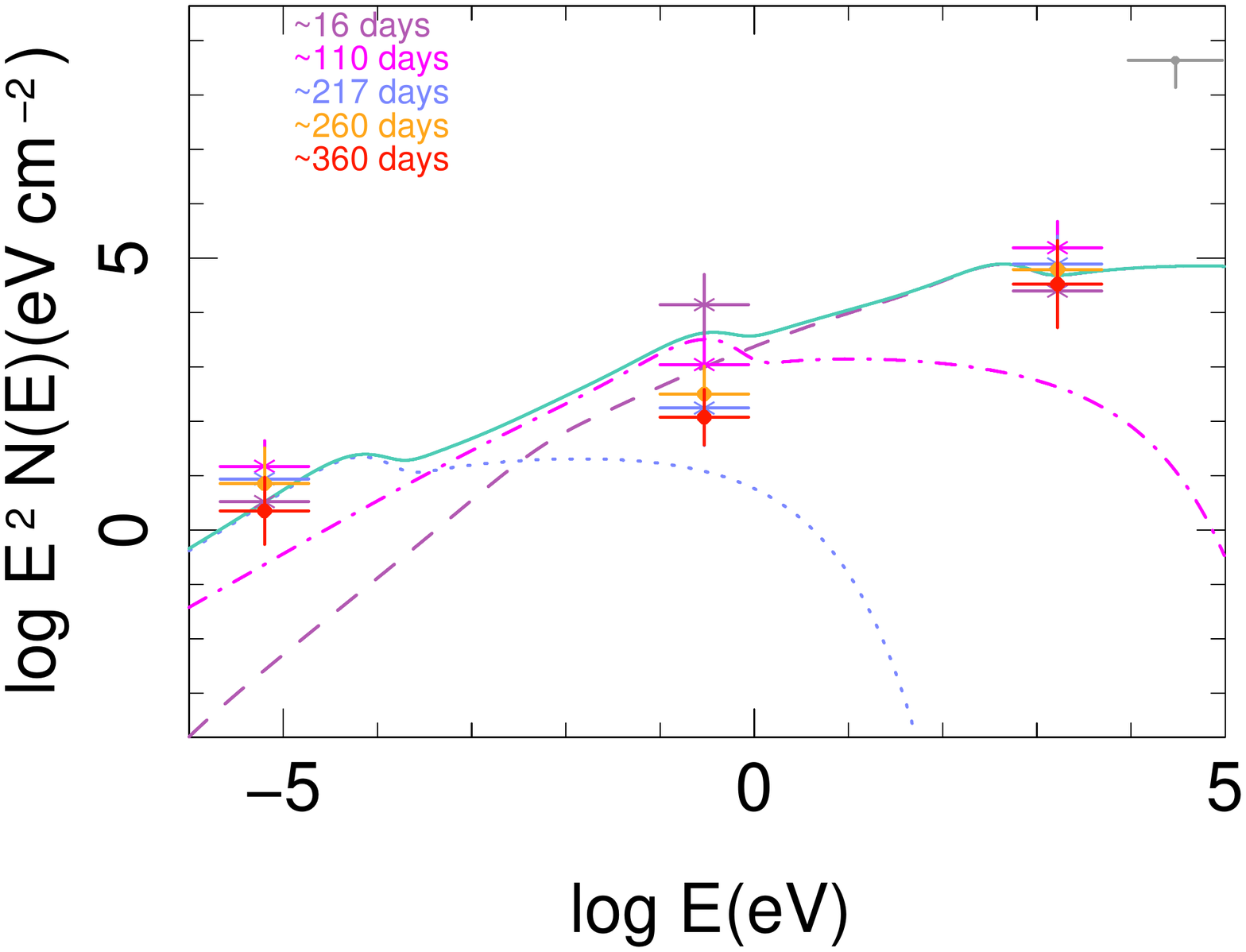} & 
\hspace{-1.5cm}\includegraphics[width=8cm]{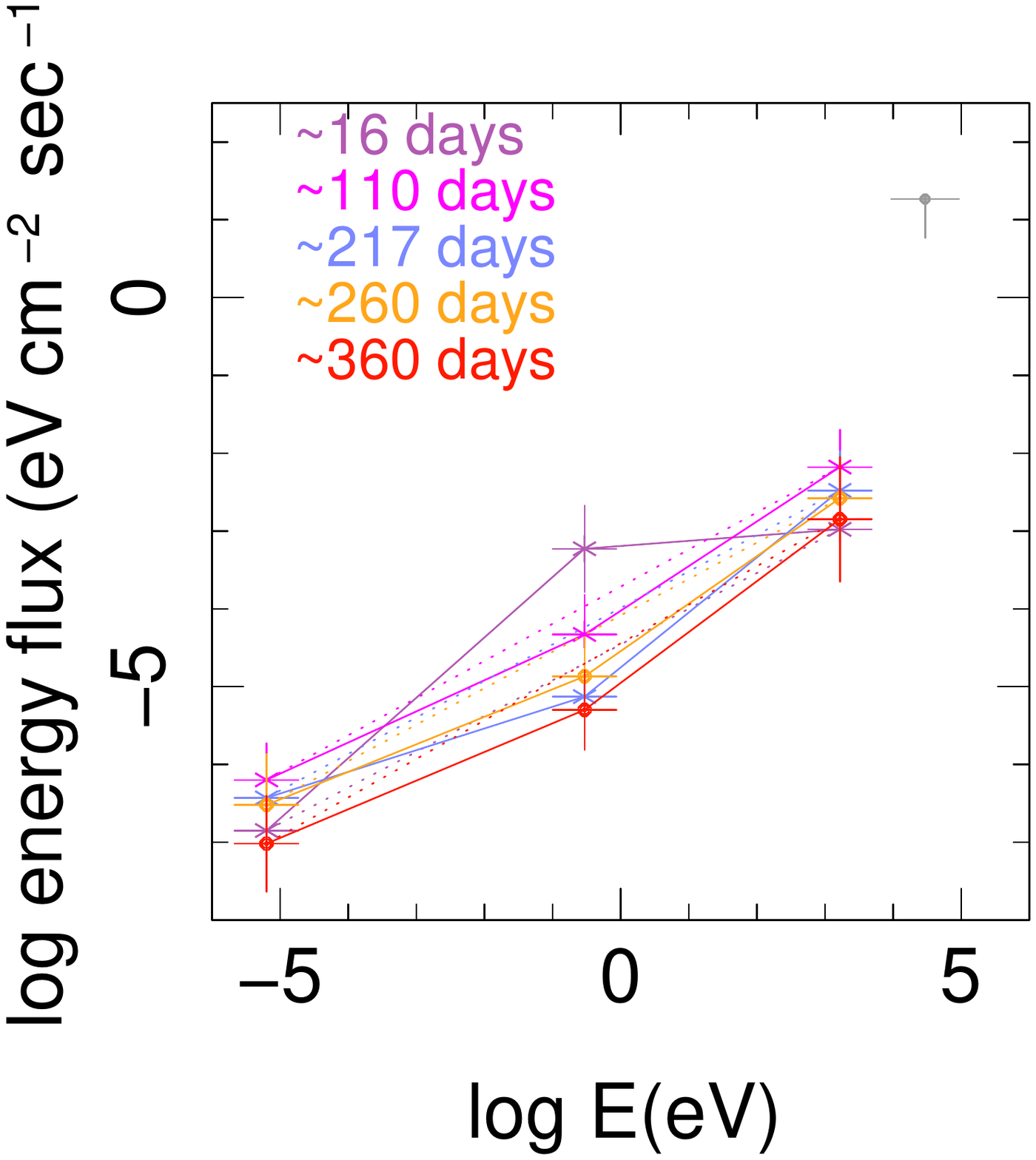} & 
\hspace{-3.5cm}\includegraphics[width=8cm]{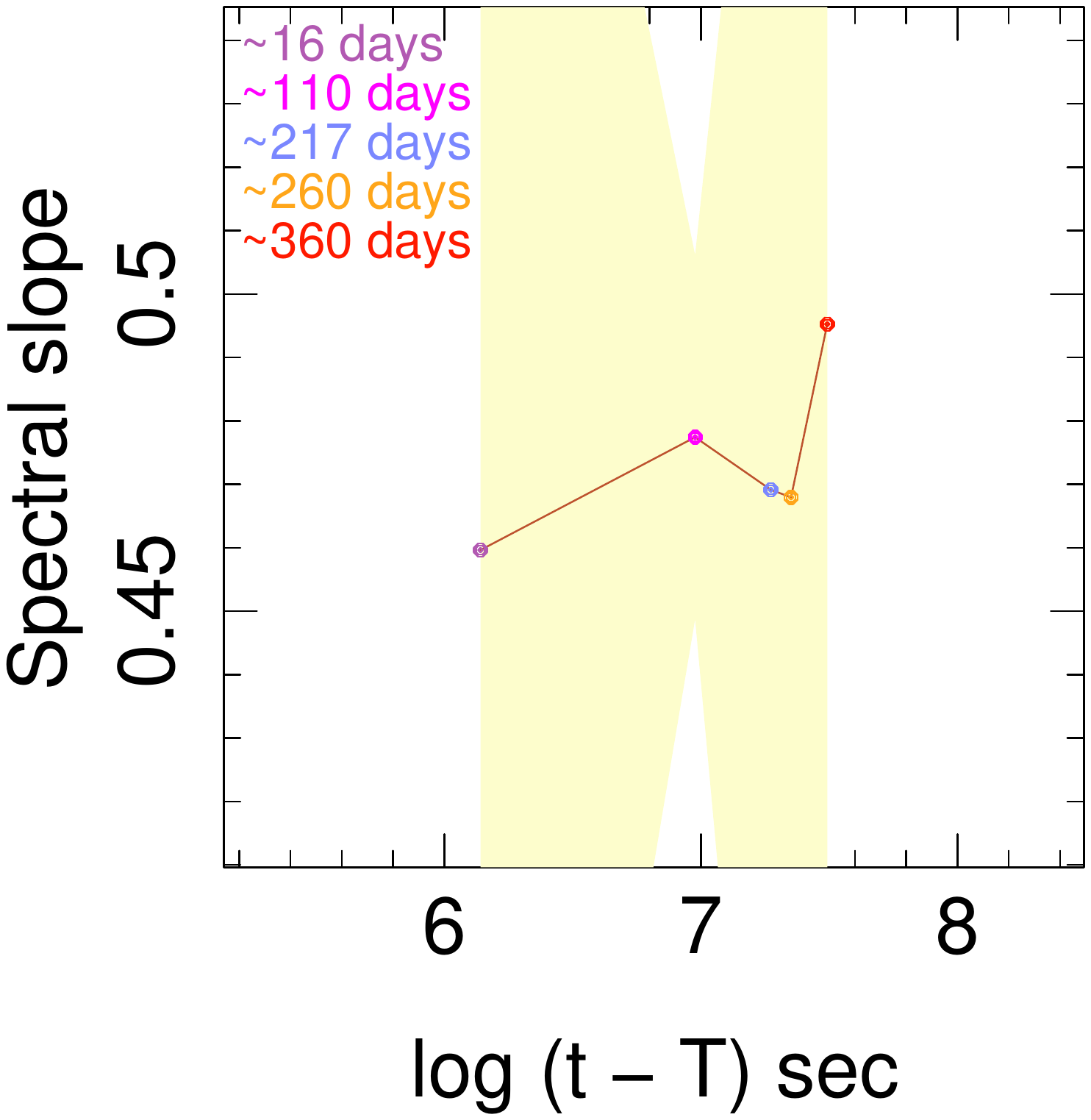} 
\end{tabular}
\caption{{\bf a)}: Spectra of components and their sum: ultra-relativistic (C1) (dash line), 
relativistic (C2) (dash-dot), mildly relativistic (C3) (dotted line), sum of 3 components 
(full line). Crosses present observations at different times in radio, optical/IR, and X-ray, 
optical and radio bands. When data for a time interval was not available an interpolation has been 
used. The width of crosses presents the width of the corresponding filter and are much larger than 
observational uncertainties. The upper limit at $E \sim 16-50$~keV is from the Swift-BAT survey 
data~\cite{batgammaupper}. To generate a pseudo-spectrum from flux measurements, we have 
normalized data such that X-ray at $T+217$~days become equal to maximum of simulated spectrum 
in the simulated energy interval. {\bf b)}: Spectrum of energy flux. The lines connecting the data 
points are added to facilitate the illustration of spectral variation. {\bf c)}: Evolution of the 
slope of pseudo-spectrum using only radio and X-ray data, that is slope of dotted lines in {\bf b)}. 
The shaded region is the estimate of uncertainty of calculated slopes according to variation rule. 
\label{fig:spectagmain}}
\end{figure}
\end{center}

Spectra of the components and total spectrum of the model are shown in Fig. \ref{fig:spectagmain}. 
They show a good consistency between simulated spectrum and the data, and as expected, amplitude of 
optical emission at $\sim T+ \mathcal{O}(10)$~days is higher than the afterglow model. Moreover, 
due to the dominance of kilonova contribution, the shape of pseudo-spectrum of energy flux shown in 
Fig. \ref{fig:spect}-b at this epoch is significantly different from those of later times. In 
addition, Fig. \ref{fig:spect}-c shows a crude broad-band spectral slope, which is determined from 
radio and X-ray data only and thereby is not contaminated by kilonova emission. Ignoring large 
uncertainties of calculated slopes, they show a behaviour similar to other GRB afterglows, namely 
softer spectrum during earliest observations around $T+\mathcal{O}(10)$~days, which gradually 
becomes harder until the peak of emissions around $T+110$~days, and finally softens at later times. 
Therefore, despite unusual brightening of afterglows, they have the same spectral trend as other GRBs.

\begin{table}
\begin{center}
\caption{Parameter set of simulated afterglow components. \label{tab:agparam}}
\end{center}
{\scriptsize
\begin{center}
\begin{tabular}{p{5mm}p{5mm}p{5mm}p{12mm}p{8mm}p{10mm}p{5mm}p{5mm}p{5mm}p{5mm}p{6mm}p{5mm}p{5mm}p{5mm}p{8mm}p{8mm}|}
\hline
Comp. & mod. & $\gamma'_0$ & $r_0$ (cm) & $\frac{\Delta r_0}{r_0}$ & $(\frac{r}{r_0})_{max}$ & $p$ & $\gamma_{cut}$ & $\kappa$ & $\delta$ & $\epsilon_B$ & $\alpha_B$ & $\epsilon_e Y_e$ & $\alpha_e$ & $N'$ (cm$^{-3}$) & $n'_c$ (cm$^{-2}$) \\
\hline 
\multirow{4}{5mm}{Ultra. rel. (C1)} 
 & 1 & 130 & $10^{16}$ & $10^{-7}$ & 1.5 & 1.8 & 100 & -0.5 & 0.5 & $0.08$ & -1 & 0.1 & -1 & $0.04$ & $5 \times 10^{22}$ \\
 & 2 & -   &   -      &  -    & 15  &  -  & 100 & 0.3 & 0.1 &  -     &  0 &  -   &  0 &   -    &    -   \\
 & 2 & -   &   -      &  -    & 20  &  -  & 100 &  0.4  & 0.05 &  -     &  1 &  -   &  1 &   -    &    -   \\
\hline
\multirow{4}{5mm}{Rel. (C2)} 
 & 1 & 5 & $10^{16}$ & $10^{-6}$ & 2 & 2.1 & 100 & -0.5 & 1 & $0.08$ & -1 & 0.1 & -1 & $0.04$ & $10^{23}$ \\
 & 2 & -   &   -      &  -    & 40  &  -  & 100 & 0.4 & 0.1 &  -     &  0 &  -   &  0 &   -    &    -   \\
 & 2 & -   &   -      &  -    & 100  &  -  & 100 &  0.5  & 1 &  -     &  1 &  -   &  1 &   -    &    -   \\
\hline
\multirow{4}{5mm}{Mildly rel. (C3)} 
 & 1 & 1.06 & $1.5 \times 10^{16}$ & $10^{-2}$ & 1.5 & 1.8 & 100 & -0.5 & 1 & $0.08$ & -1 & 0.02 & -1 & $0.008$ & $10^{24}$ \\
 & 2 & -   &   -      &  -    & 10  &  -  & 100 & 0. & 0.1 &  -     &  0 &  -   &  0 &   -    &    -   \\
 & 2 & -   &   -      &  -    & 10  &  -  & 100 &  1  & 1 &  -     &  1 &  -   &  1 &   -    &    -   \\
\hline
\end{tabular}
\end{center}
}
\begin{description}
\item {$\star$} Each data line corresponds to one simulated regime, during which quantities 
listed here remain constant or evolve dynamically according to fixed rules. A full simulation 
of a burst usually includes multiple regimes (at least two). 
\item {$\star$} Horizontal black lines separate time intervals (regimes) of independent 
simulations identified by the label shown in the first column.
\item {$\star$} A dash as value for a parameter presents one of the following cases: it is 
irrelevant for the model; it is evolved from its initial value according to an evolution 
equations described in~\cite{hourigrb,hourigrbmag}; it is kept constant during all regimes.
\end{description}
\end{table}
\subsubsection{Degeneracies} \label{sec:degen}
\begin{center}
\begin{figure}
\begin{tabular}{p{7cm}p{7cm}}
a) \includegraphics[width=7cm]{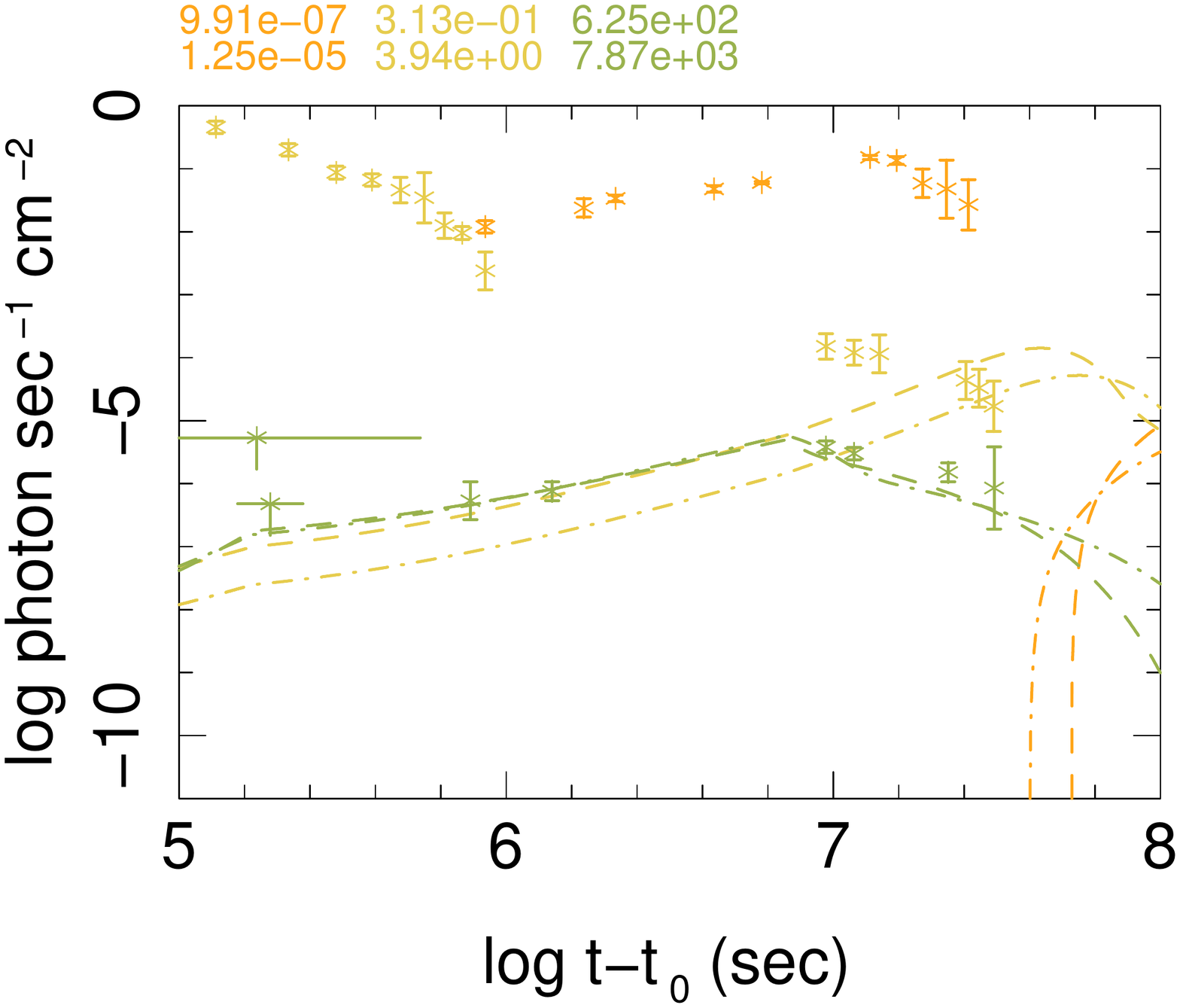} & 
b) \includegraphics[width=7cm]{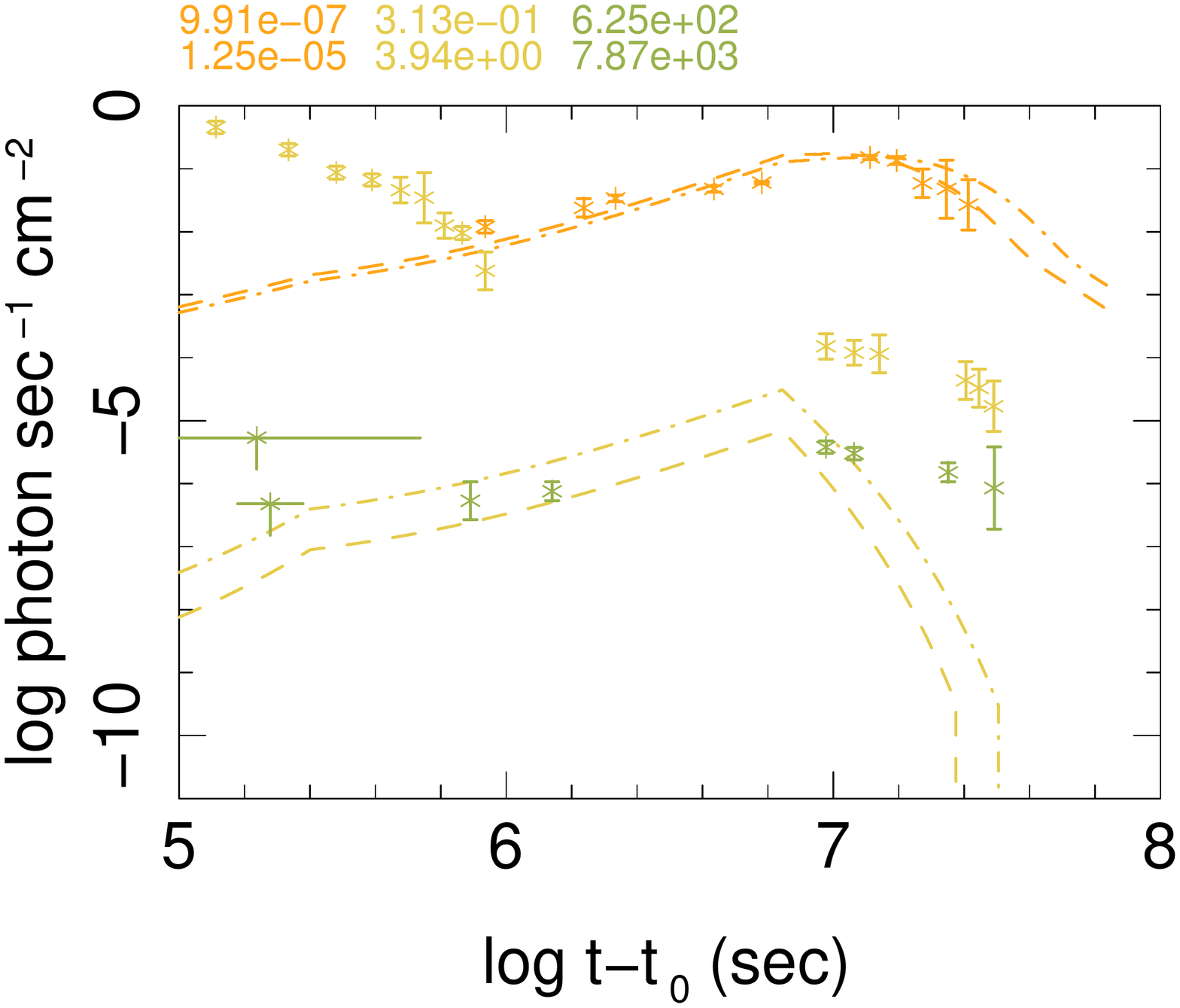} \\
c) \includegraphics[width=7cm]{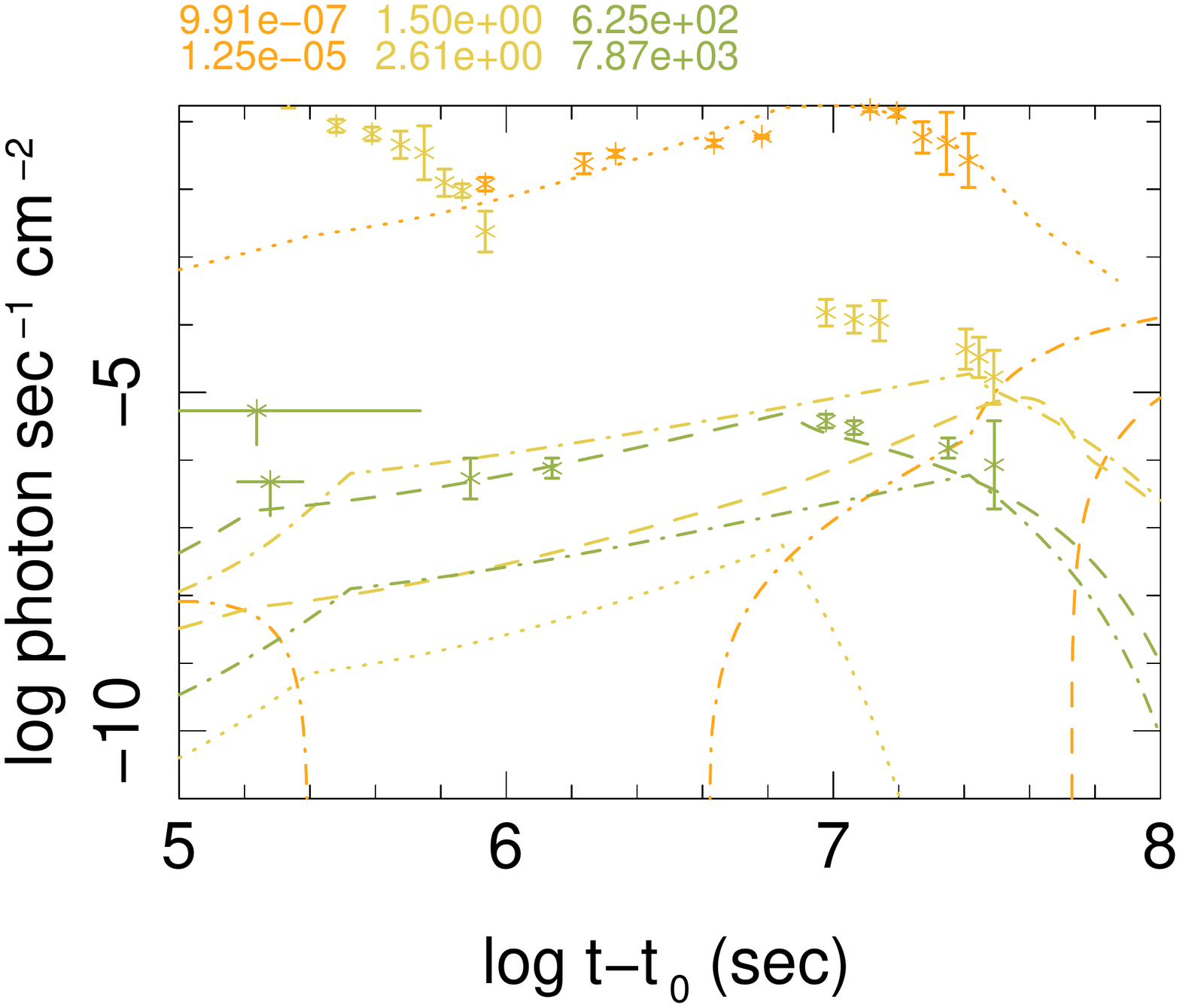} & 
d) \includegraphics[width=7cm]{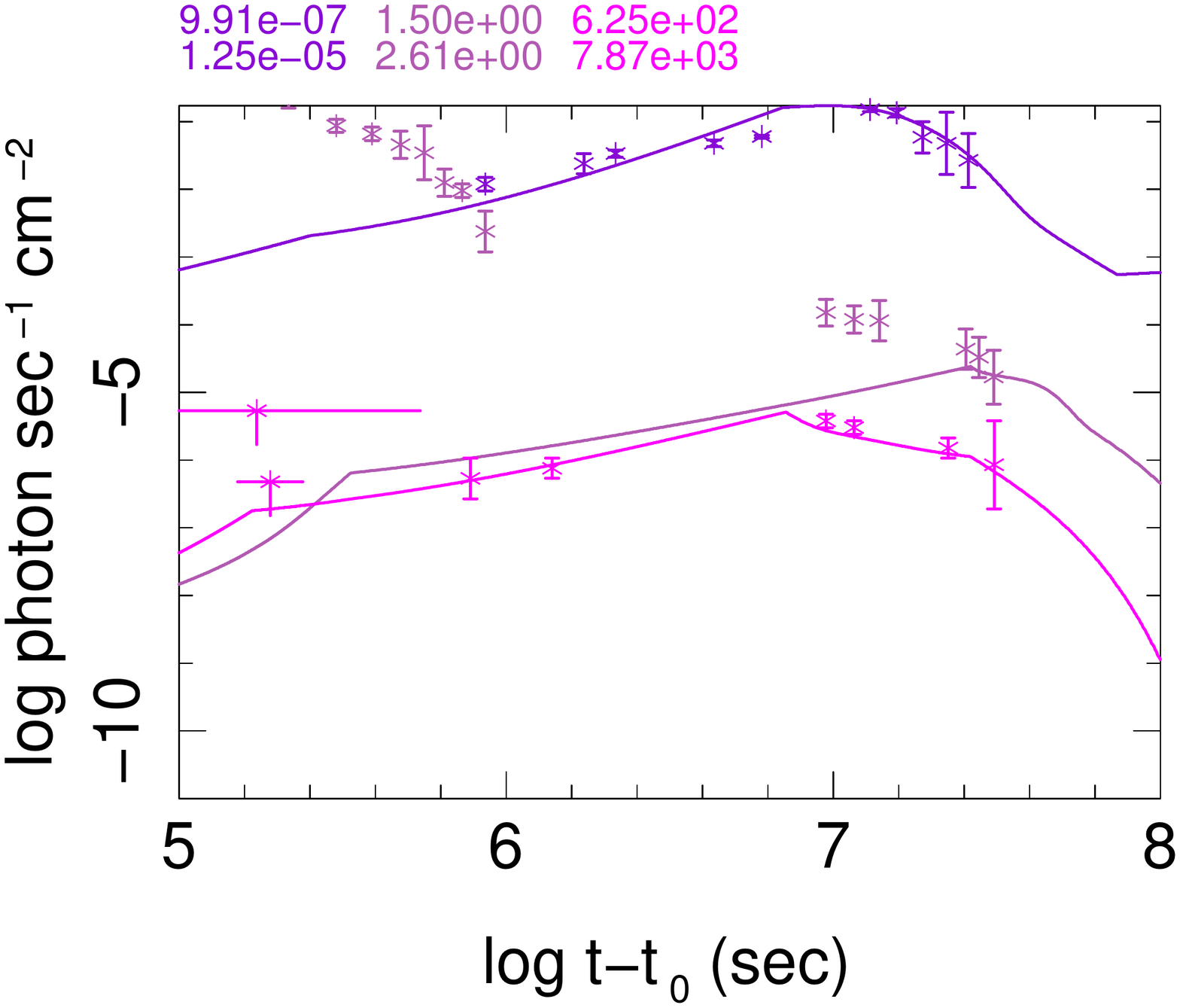}
\end{tabular}
\caption{Variants of components of the model of Table \ref{tab:agparam} which fit the data: 
a) A variant of component C1 with $\Delta r_0/r_0 = 10^{-6}$, $N' = 0.004$~cm$^{-3}$ (dash-dot); 
b) A variant of C3 with $\gamma'_0 = \Gamma = 4$, $r_0 = 10^{16}$~cm, $\Delta r_0/r_0 = 10^{-3}$, 
$N' = 0.001$~cm$^{-3}$, $n'_c = 10^{25}$~cm$^{-2}$ (dash-dot); 
c) Variant of C2 component with $\gamma'_0 = \Gamma = 30$, $\Delta r_0/r_0 = 10^{-5}$, 
$n'_c = 10^{22}$~cm$^{-2}$, $N' = 0.008$~cm$^{-3}$ (dash-dot); In all these plots dashed lines 
correspond to the components described in Table Table \ref{tab:agparam}; d) Total light curves for 
a model with C1 and C2 as in Table \ref{tab:agparam} and C2 as shown in c). \label{fig:variants}}
\end{figure}
\end{center}

Despite the fact that the model described in Table \ref{tab:agparam} provides a good fit to the 
data, some issues should be considered before making any conclusion. Notably, as we explained 
in Sec. \ref{sec:model} parameters of the phenomenological formalism 
of~\cite{hourigrb,hourigrbmag} are not completely independent, and thereby they may 
be degenerate. Thus, before interpreting this model and concluding properties of 
the jet from it, we must consider implications of these degeneracies.
 For this purpose, change of the characteristics of the model as a function of their 
variation is studied in some extent in~\cite{hourigw170817lateagjet}. Here we summarize 
them by showing in Fig. \ref{fig:variants} a few variants for components of the model which 
can fit the data as good as those presented in Table \ref{tab:agparam}. Specifically, 
Fig. \ref{fig:variants}-a shows an alternative to component C1 with smaller ISM/circum-burst 
density but longer jet extent. It fits X-ray data as good as C1 in Table \ref{tab:agparam}, 
but future observations should be able to distinguish between them. 

Variant of C3 shown in Fig. \ref{fig:variants}-b is important because it has a Lorentz factor consistent 
with superluminal motion of radio counterpart~\cite{gw170817lateradiosuprlum,gw170817lateradiosuprlum0}. 
Alternatively, with slight modification of column densities and/or thicknesses of active regions 
$\Delta r$ in these models we can consider both of them as components of the full model. Indeed, 
evidence for a mildly relativistic component with small Lorentz factor - {\it a cocoon} - is observed 
in long GRB~171205A and its associated supernova SN2017uk~\cite{grb171205asn2017uk}, 
and has $\beta \sim 0.3$, corresponding to $\Gamma \sim 1.05$, i.e. similar to C3 component in 
Table \ref{tab:agparam}. Therefore, model C3 and its variant in Fig. \ref{fig:variants}-b may present 
sectors further and closer to the jet axis, respectively.

Another important conclusion from the study of parameter degeneracy is the influence of distance 
between merger and location of external shocks on the slope of the light curve and time of its turnover.
See~\cite{hourigw170817lateagjet} for example of models with shorter distances than 
$\sim 10^{16}$~cm used in the model of Table \ref{tab:agparam}. The peak light curves in these 
models are too early and inconsistent with the data.

\subsection {Other models and their consistency with data} \label{sec:othermodel}
In our knowledge~\cite{hourigw170817} is the only detailed modeling and analysis of the prompt 
gamma-ray emission of GW/GRB~170817A. Therefore, here we only compare afterglow models 
described in Sec. \ref{sec:ag} with some of the proposed models in the literature.

Fitting afterglow data with a multi-component model in~\cite{hourigw170817lateagjet} is not 
unique and some other authors have similarly modelled late afterglows of GW/GRB~170817A in 
this manner. For instance,~\cite{gw170817latexary,gw170817latexoptradio1} consider 
a two component structured jet model with a top-hat ultra-relativistic component and 
$\Gamma \sim 100$ in its inner $\theta \lesssim 9^\circ$, where $\theta$ is angle with 
respect to symmetry axis of the outflow, and a component with a decreasing Lorentz factor 
in the angular interval $10^\circ \lesssim \theta \lesssim 60^\circ $ with a mean 
$\Gamma \sim 10$. They use a relativistic jet simulation code to find that a line of sight 
angle from the jet axis $\theta_v \sim 20^\circ$. This model is very similar to the model 
described in the latest version of~\cite{gw170817latexraystructjet}, but Lorentz factor 
and energy profile of the jet in the two works are different. Authors 
of~\cite{gw170817grbonaxis} also conclude an initially ultra-relativistic jet from their 
analysis of afterglows. Authors of~\cite{gw170817lateopthstir} consider two profiles 
for the jet, one similar to 2-component model 
of~\cite{gw170817latexary,gw170817latexoptradio1} and the other with Gaussian 
energy and Lorentz factor profiles. Both models find a small central core opening angle of 
$\lesssim 5^\circ$. In their 2-component model high and low Lorentz factors are $\sim 100$ and 
$\sim 5$, respectively. However, in the Gaussian model on-axis Lorentz factor can be as large as 
$900$. A phenomenological function which effectively has two coupled components is used by 
\cite{gw170817lateradio300}.

It should be pointed out that asymptotic formulation of synchrotron emission from external shock 
formulation in~\cite{emission1}, which is used in all the cited works, considers a uniform 
spherical ejecta. Therefore, conclusions about viewing angle of the jet in the cited works is 
based on the value of Lorentz factor and beaming of emissions from a relativistic 
source~\cite{emissionbook}. An interpretation of the multi-component model of 
Table \ref{tab:agparam} in the framework of~\cite{emission1} formulation can be found 
in~\cite{hourigw170817lateagjet}.

\section {Interpretation of models} \label{sec:interpret}
In this section we use models of prompt and afterglows described in the previous section to check 
their consistency, remove degeneracies, and correlate them to properties of progenitor neutron 
stars, their environment and merger, and its ejecta.

\subsection{Selecting between prompt models}
Although simulations of prompt gamma-ray emission show that both an ultra-relativistic jet with a 
Lorentz factor of $\mathcal{O}(100)$ and a relativistic jet with a Lorentz factor of 
$\mathcal{O}(10)$ are consistent with data, afterglows rule out the latter case. We should remind 
that the analysis of prompt gamma-ray emission in~\cite{hourigw170817} was performed well before 
the relatively early turn over of afterglows. Therefore, they could not be used to distinguish 
between the two possible range of initial Lorentz factor, which might additionally discriminate 
between a significantly off-axis view of the jet and otherwise. For this reason, 
in~\cite{hourigw170817} several other arguments were given in favour of an ultra-relativistic jet 
with $\Gamma \sim \mathcal{O}(100)$. Here we briefly review them because this approach may be 
useful for analysing future GW/GRB events.

\begin{figure}
\begin{center}
\begin{tabular}{p{6cm}p{6cm}}
\includegraphics[width=6cm]{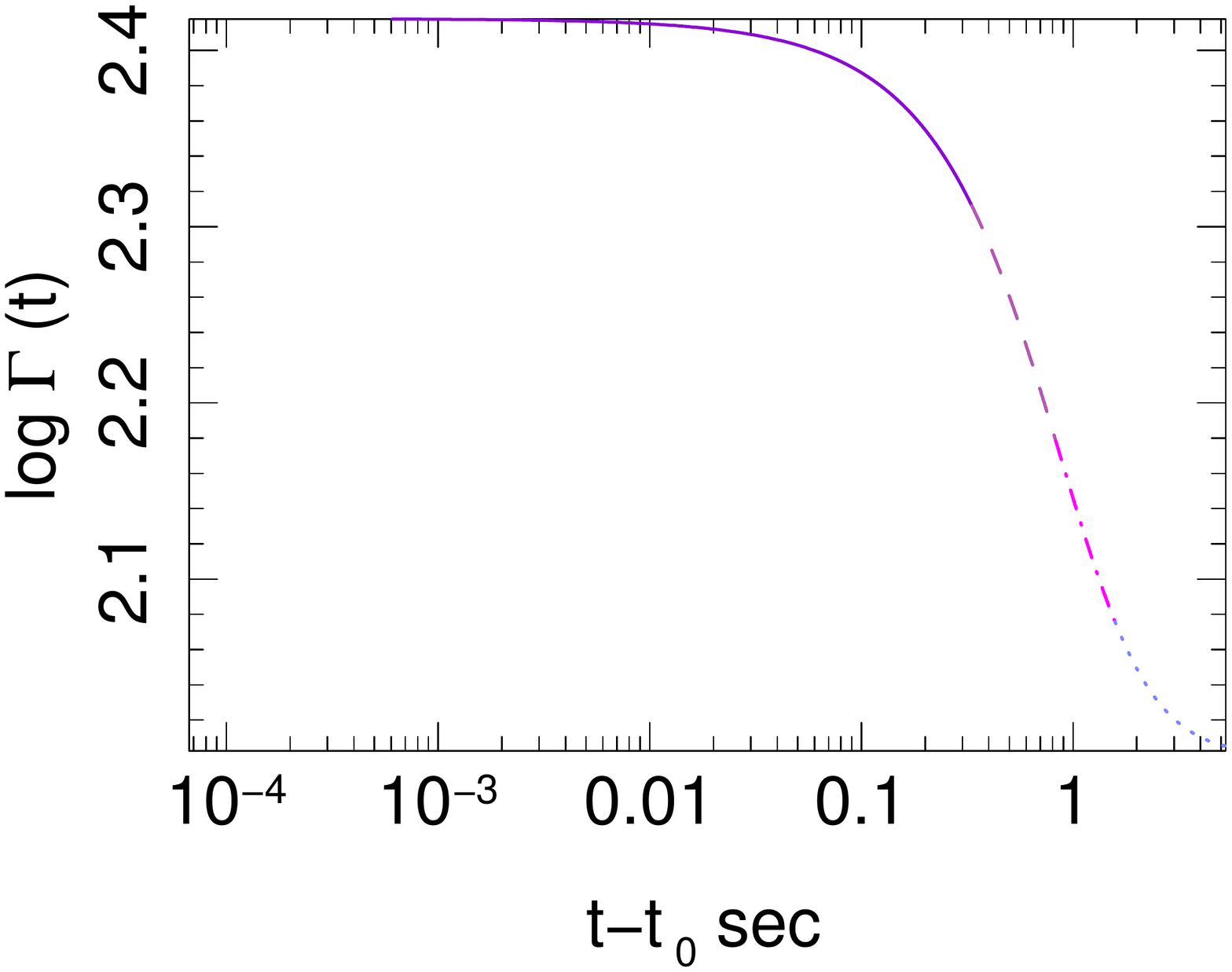} & 
\includegraphics[width=6cm]{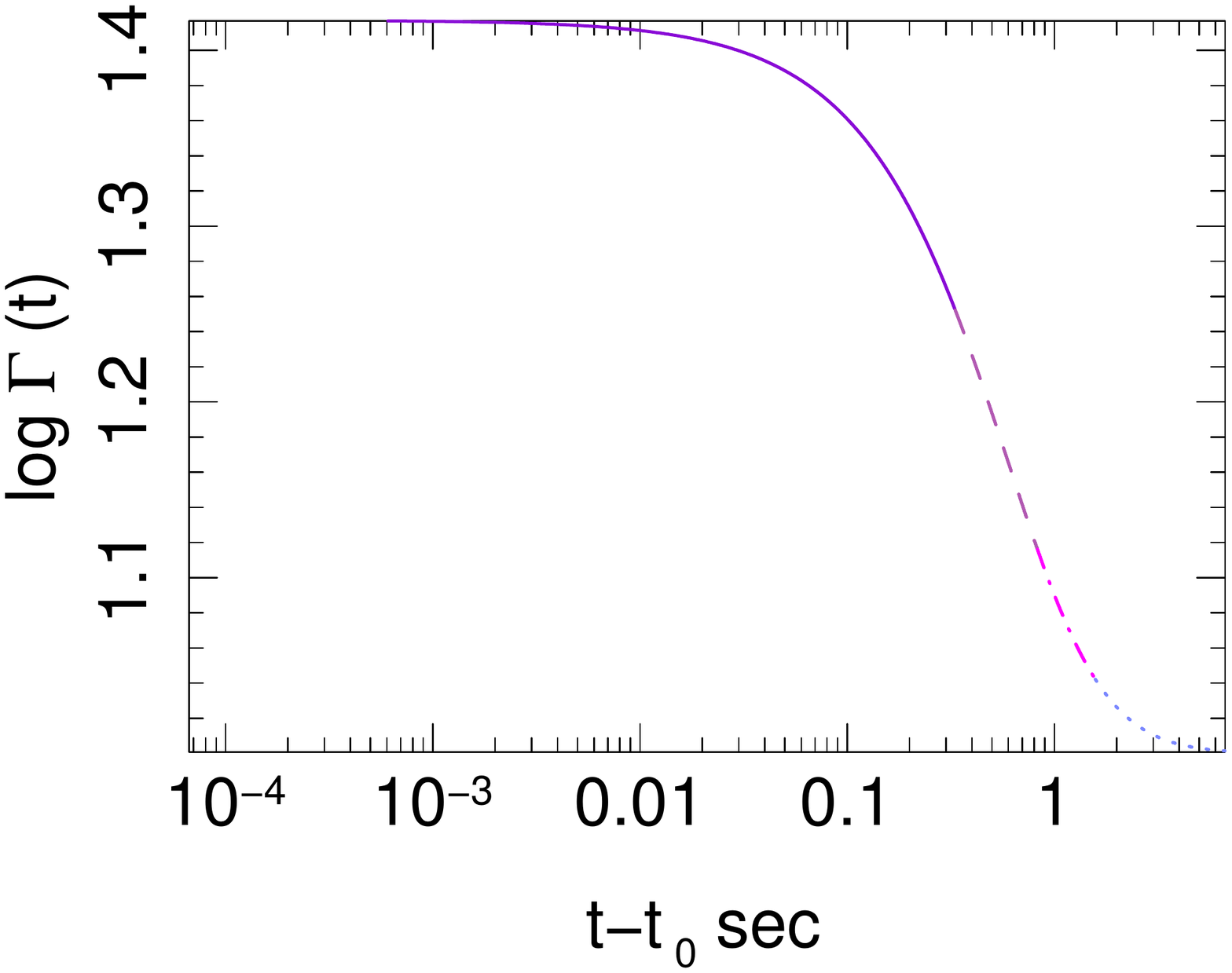}
\end{tabular}
\end{center}
\caption{Evolution of Lorentz factor of the fast shell with respect to a far observer at the 
redshift of the source during internal shocks. Left: ultra-relativistic jet (Model 1 in 
Table \ref{tab:param}); Right: relativistic jet (Model 2 in Table \ref{tab:param}). Different line 
styles present simulation regimes. \label{fig:promptgamma}}
\end{figure}

Giving relation between emitted and received power from a relativistic source 
$(dP_e/d\omega d\Omega)/(dP_r/d\omega d\Omega) = 1/\Gamma^2 (1+\beta \cos \theta_v)$, it is clear 
that off-axis view alone cannot explain intrinsic faintness of the burst, if the jet is uniform. 
Therefore, GW/GRB~170817 had to have a structured jet. On the other hand, even during prompt 
internal shocks energy dissipation significantly reduces Lorentz factor, see 
Fig. \ref{fig:promptgamma}, and scattering of particles by induced electromagnetic fields 
generates a lateral expansion. This process should bring dissipated material and its emission 
to the line of sight~\cite{grboffaxis} and should have been detectable as a tail emission, which 
is detected in some short bursts, and/or it should have generated bright early afterglows. None 
of these emissions are observed in GW/GRB 170817A event. Thus, according to this argument even 
an off-axis view of a structured jet cannot explain the faintness of the prompt gamma-ray and 
early afterglows and their late and slow brightening.

Another criteria for choosing between low and high Lorentz factor candidate models of 
Table \ref{tab:param} is plausibility of parameters which characterize shocks and synchrotron 
emission. Table \ref{tab:param} shows that in the moderately relativistic Model 2 smaller Lorentz 
factor is compensated by a higher fraction of energy transferred to electrons (more generally charged 
leptons), which is $\sim 3$ times larger than in the ultra-relativistic jet Model 1. However, 
apriori due to the low density of the jet in this model interaction between charged particles and 
their scattering had to be rarer and induced electric and magnetic fields weaker. Moreover, low 
electron yield of neutron rich BNS 
ejecta~\cite{nsmergerrprocsimul,nsmergerrprocsimulhres,nsmergerrprocsimul0} should have made the 
transfer of kinetic energy to electrons even harder. Estimation of electron yield 
$Y_e$ for various components of the ejecta of GW~170817 event based on the observation of r-process 
products~\cite{nstarmergeejecta,gw170817rprocessth} are: $Y_e \sim 0.1 - 0.4$ for dynamical 
component, $Y_e \sim 0.3$ for wind, and $Y_e \sim 0.25$ in another wind 
component~\cite{gw170817rprocess}. Considering the value of $\epsilon_e Y_e$ in the low Lorentz 
factor Model 2, the effective fraction of kinetic energy transferred to electrons had to be 
$\epsilon_e \sim 0.1-0.3$, which is much higher than $\epsilon_e \lesssim 0.1$ found in Particle In 
Cell (PIC) simulations~\cite{fermiaccspec,fermiaccspec0,fermiaccspec1}. By contrast $\epsilon_e$ of 
Model 1 is comfortably in the PIC range. 

In conclusion, Model 1 seems more plausible than Model 2. Considering parameters of Model 1 as 
approximately presenting properties of the jet before prompt internal shocks, according to 
simulations of~\cite{hourigrbmag} its bulk Lorentz factor in our direction was a few folds smaller 
than typical short bursts. Additionally, densities of colliding shells were more than one order 
of magnitude less brighter GRBs. Moreover, these findings demonstrate that most probably in 
addition to off-axis, intrinsic properties of progenitor neutron stars and dynamics of their 
merger and ejecta were responsible for the faintness of GRB~170817A. In addition, the lack of 
bright short GRBs at low redshifts in Fig. \ref{fig:allsgrb} is an evidence for the influence of 
progenitor BNS evolution on the strength of GRBs produced during their merge.

\subsection{Kinematic of the jet at late times}
The structure of 3-component jet model at late times, that is $>T+10$~days, discussed in 
Sec. \ref{sec:ag} is in agreement with our arguments in favour of an ultra-relativistic prompt jet 
with a Lorentz factor $\Gamma \sim 100$ at the end of internal shocks. As discussed earlier, 
exploration of the parameter space of external shocks 
in~\cite{hourigw170817ag,hourigw170817lateagjet} to find a model with only relativistic or mildly 
relativistic Lorentz factor was not successful. Apriori Lorentz factor of the jet at the location 
of external shocks $r_e$ should be smaller than its value at the end of internal shocks at $r_i$ 
because weaker internal shocks and cooling of shocked material should reduce its kinetic energy.
Therefore, the presence of an ultra-relativistic component with roughly the same Lorentz factor 
in the afterglow model means that despite all odds, a fraction of ultra-relativistic jet had 
survived up to long distances. However, comparison of the column density of the prompt jet in 
Table \ref{tab:param} with that of component C1 in Table \ref{tab:agparam} show that its column 
density at the time of external shocks was reduced by a factor of $\sim 200$. If the jet were an 
adiabatically expanding cone, its column density had to decline by a factor of 
$(r_i/r_e)^{-2} \sim 10^{12}$. The much smaller dilation factor according to models described here 
means that the material inside the jet had an internal coherence and collimation - most probably 
through imprinted electric and magnetic fields in the plasma. Thus, we conclude that its geometry 
and expansion were closer to a boosted cylinder rather than an adiabatic cone. In addition, 
in~\cite{hourigw170817lateagjet} it is shown that non-adiabatic expansion of jet cannot be associate 
to accretion of material, which could completely extinguish its boost.

\subsubsection{Late time jet structure and interpretation of components} \label{sec:offdark}
As discussed earlier, unusual characteristics of GRB~170817A may be in some extent due to the 
off-axis view of the jet. Indeed, observation of gravitational waves from this transient indicates 
an orbital inclination angle of 
$ 18^\circ \lesssim \theta_{in} \lesssim 27^\circ$~\cite{gw170817decline}. Moreover, superluminal 
motion of radio afterglow with an apparent speed of 
$\beta_{app} = 4.1 \pm 0.5$~\cite{gw170817lateradiosuprlum,gw170817lateradiosuprlum0} is evidence 
of an oblique view of its source.

Due to relativistic beaming of photons, off-axis view of the jet has significant consequences for 
observations. A far observer receives synchrotron emission only from a cone with half angle 
$\theta_{max} \equiv \arcsin (1/\Gamma)$ with respect to the line of sight. For components C1, C2, 
and C3 of the afterglow model listed in Table \ref{tab:agparam} these angles are 
$\theta_{max_1} \sim 0.5^\circ$, $\theta_{max_2} \sim 11.5^\circ$ and $\theta_{max_3} \sim 65^\circ$ or 
$14.5^\circ$ for C3 in Table \ref{tab:agparam} and the model in Fig. \ref{fig:variants}-b as its 
alternative or an additional component, respectively. This leads us to conclude that 
the 3 (or 4)-component model of the jet at the time of external shocks presents a structured jet 
and components of the model approximately present its angular structure and characteristics of its 
shock on the surrounding material from our line of sight up to its outer boundary\footnote{For the 
time being, the simulation code used in this work uses an analytical expression 
for determining synchrotron/self-Compton flux. In addition, terms depending on higher order of 
angle between emitting element and line of sight $\theta$ are neglected. This is a good approximation 
when $\Gamma \gg 1$. Under this simplification $\theta$ dependence is only through a 
$(cos (\theta) + \beta) \leq (cos (\theta) + 1)$ factor, which must be integrated between 
$\theta_0 \geq -\theta_{max}$ and $\theta_1 \leq \theta_{max}$. However, angular size of emitting 
surface may be smaller than $2 \theta_{max}$. In this case, integration over maximum visible angle 
over-estimate the flux. But the difference would be at most a factor of few and comparable to other 
uncertainties of the model. Indeed, for this and other simplifications and approximations applied 
to the model and its simulations that we should consider parameters as order of magnitude 
estimations.}. In particular, their Lorentz factors and column densities present azimuthal variation 
of material load of the polar outflow and its velocity up to a $\cos \theta$ factor, that is 
$\Gamma_{simul,i} = \Gamma_i (\theta_i) \cos \theta_i$, where $\theta_i$ is angle between centroid of 
component $i$ of the model and the line of sight. 

\begin{figure}
\begin{center}
\includegraphics[width=9cm]{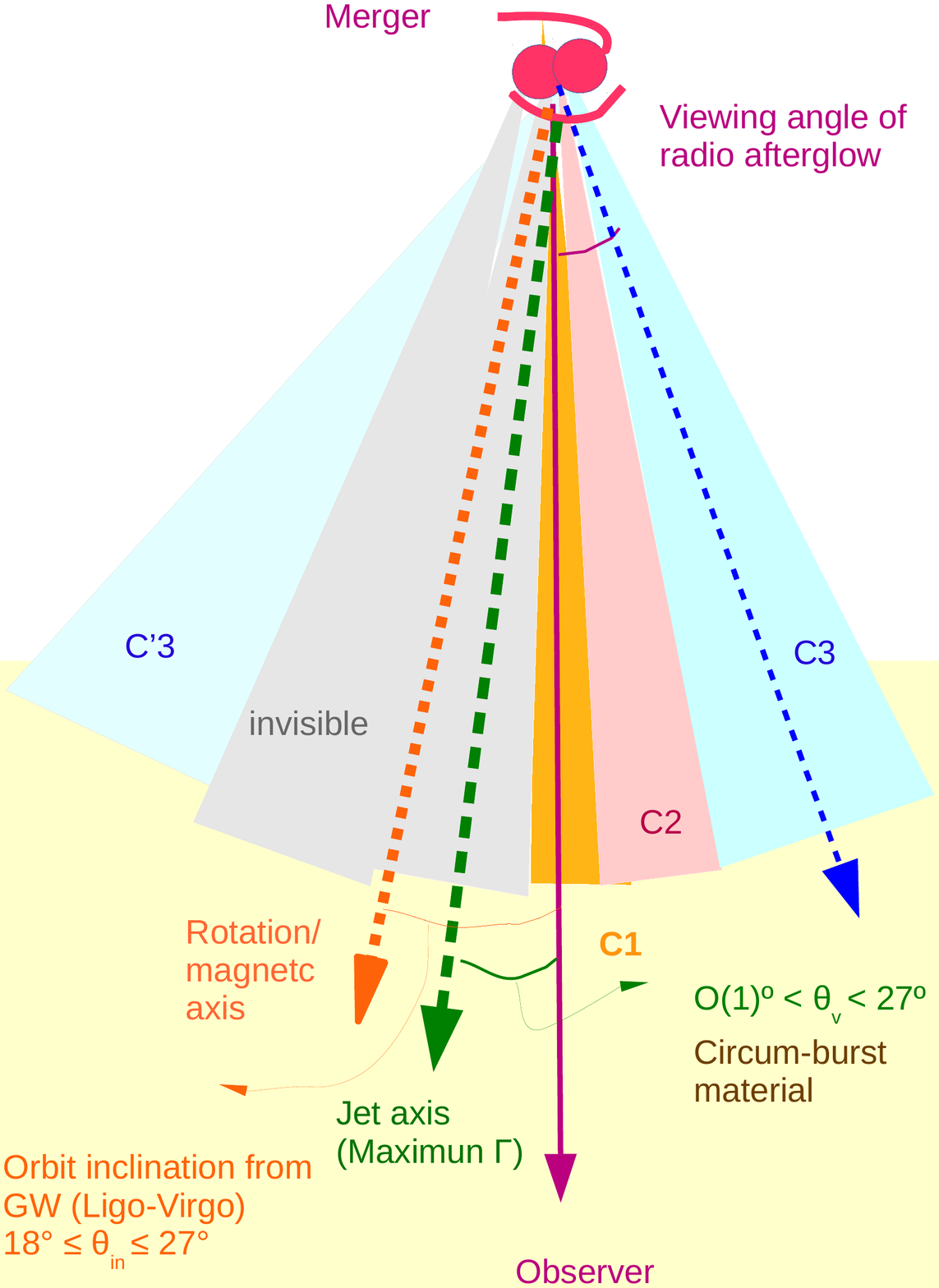}
\caption{Schematic description of polar outflow of merger at the time of its encounter with 
circum-burst material. C1, C2, C3 refer to components of the simulated model. Grey shaded 
region on the opposite side of the jet with respect to observer's line of sight is approximately 
invisible because of its large Lorentz factor and off-axis angle. Nonetheless, C'3 region 
which has even larger off-axis may be visible if its Lorentz factor is sufficiently low. Therefore, 
there can be a contribution to component C3 of the model from emission of this region. 
In any case, due to its large off-axis contribution of C3' would be subdominant. For the 
sake of simplicity here we have assumed that magnetic field direction and rotation axis 
coincide. This may not be true. \label{fig:angles}}
\end{center}
\end{figure}
Fig. \ref{fig:angles} shows a schematic presentation of this interpretation, components of the 
model, and their positions with respect to our line of sight. We remind that jet axis may be 
inclined with respect to rotation axis of the merger~\cite{grbjetsimul}. Considering this 
possibility and estimated inclination of orbit, viewing angle is constraint to 
$\mathcal{O}(1^\circ) \lesssim \theta_v \lesssim 27^\circ$. This is consistent with estimations 
of~\cite{gw170817latexary,gw170817latexoptradio1,gw170817lateradiosuprlum}, but not very 
restrictive. Moreover, it does not provide any information about Lorentz factor of the invisible 
core of the jet. Additionally, the maximum off-axis of components according to their Lorentz 
factor does not fix their centroid. To estimate these quantities we assume a Gaussian profile for 
the jet and apply constraints on the centroid angles and on the parameters of the profile, 
see~\cite{hourigw170817lateagjet} for details. From this analytical approximation we find: 
$\theta_2 \sim 9^\circ-11^\circ$, $\theta_3 \sim 12^\circ-15^\circ$, assuming on-axis Lorentz factor 
$\Gamma_{max} < 1000$. In~\cite{hourigw170817lateagjet} it is shown that the range of allowed values 
for the viewing angle $\theta_v$ is strongly correlated with $\Gamma_{max}$ and is restricted to 
$5^\circ \lesssim \theta_v \lesssim 7^\circ$ for $\Gamma_{max} \sim 250$ and to 
$14^\circ \lesssim \theta_v \lesssim 18^\circ$ for $\Gamma_{max} \sim 1000$. For alternative C3 model 
in Fig. \ref{fig:variants}-b the range of allowed angles is even more 
restricted: $\theta_v \sim 7.5^\circ-8^\circ$, for $\Gamma_{max} \sim 250$ and 
$\theta_v \sim 8^\circ- 15^\circ$, for $\Gamma_{max} \lesssim 1000$, and $\theta_2 \sim 10.5^\circ$, 
$\theta_3 \sim 11.5^\circ$, for any standard deviation of the Gaussian profile as long as 
$\Gamma_{max} < 1000$. Marginalizing over other parameters, the angle between jet axis and rotation 
axis is constraint to $\sim [7^\circ-15^\circ]$. An exponential profile does not lead to acceptable 
values for the parameters and is ruled out. 

\begin{figure}
\begin{center}
\begin{tabular}{p{5cm}p{5cm}}
\includegraphics[width=10cm]{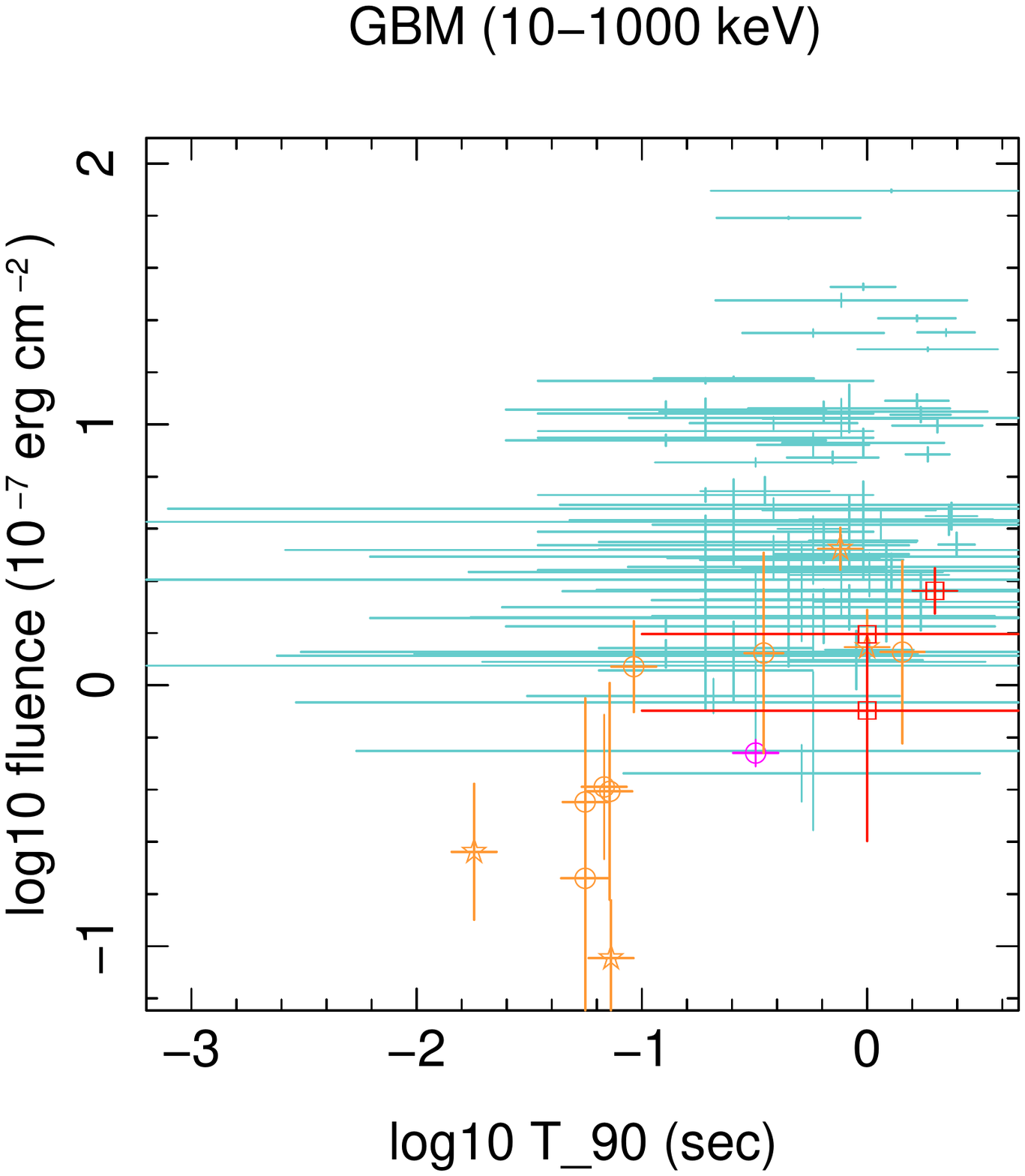} & 
\hspace{-1.5cm}\includegraphics[width=7cm,angle=90]{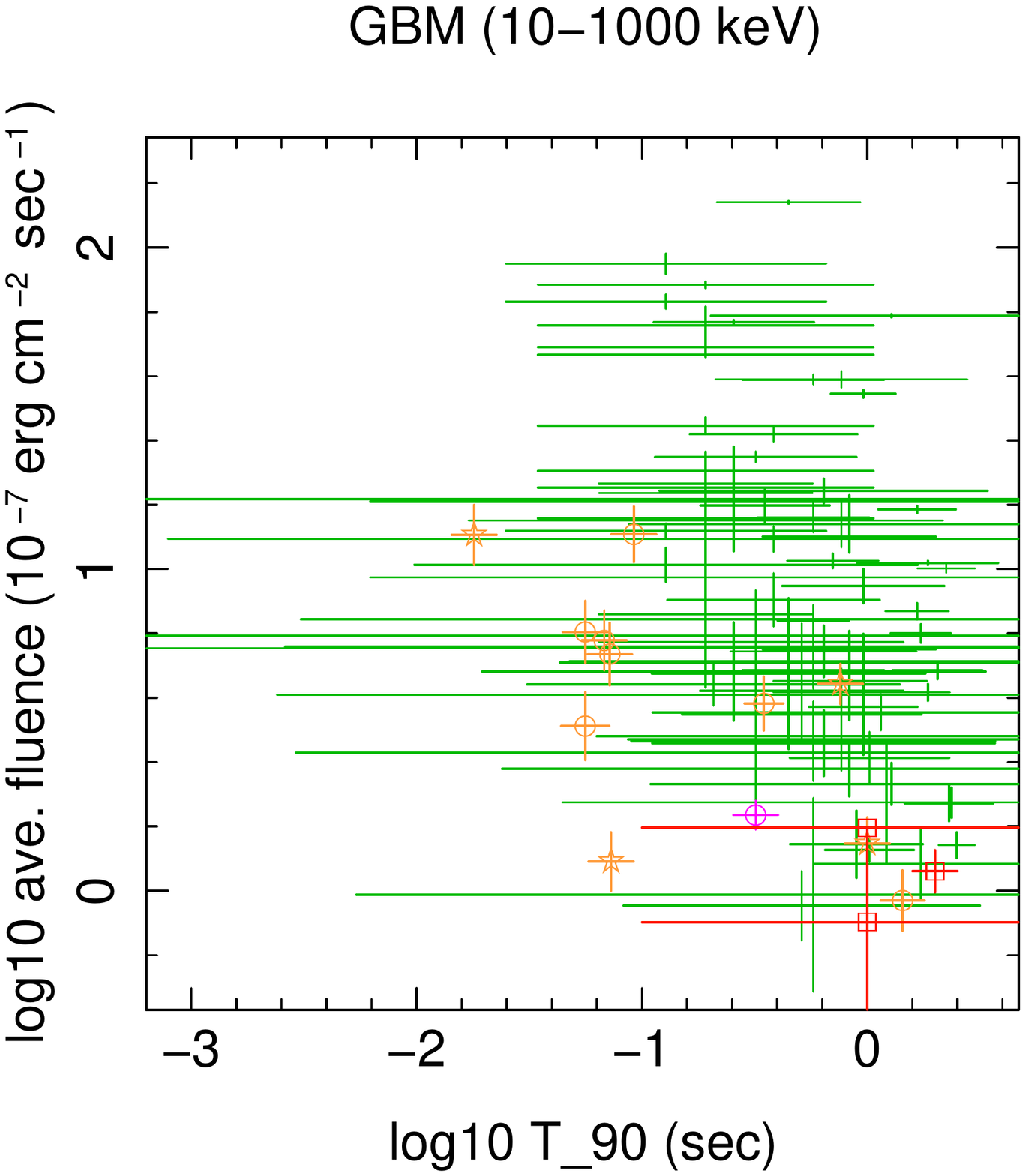}
\end{tabular}
\caption{Fluence and flux of nearby Swift short GRBs, GW-BNS events and Swift X-ray dark short GRBs. 
Left: Fluence in $15-350$~keV band; Right: Average flux. Symbols: Nearby $z < 0.2$ short GRBs 
(star), X-ray dark (circle), GW events (square), Fermi-GBM short GRBs (data without symbol). 
GRB~080121 is distinguished by its color/grey level from other X-ray dark GRBs. For Ligo/Virgo 
190425z and Ligo/Virgo 190510g only upper limit on gamma-ray is available and a $T_{90} = 1$~sec 
is used to calculate average flux. \label{fig:darkgrb}}
\end{center}
\end{figure}

Although above constraints are more restrictive than those only based on the orbit inclination, 
due to degeneracies in the parameter space it is still impossible to judge whether the core of 
the jet had a typical Lorentz factor of $\sim 500$~\cite{hourigrbmag} or had a smaller boost. 
The latter case means that off-axis viewing angle was only partially responsible for unusual 
properties of GRB~170817A. We remind that the probability of viewing a jet exactly on axis is 
very small and a deviation is always expected. Moreover, prompt gamma-ray emission of GRB~170817A 
is somehow similar to dark short bursts - those without an X-ray counterpart at 
$\lesssim T+ \mathcal{O}(10)$~ksec, see Fig. \ref{fig:darkgrb}. Although some of these GRB 
candidates may be faint flares from Soft Gamma-ray Repeaters (SGR) in nearby galaxies, others - 
specially those with longer $T_{90}$ - might be related to BNS mergers similar to GW/GRB~170817A and 
two other BNS merger gravitational wave candidate events Ligo/Virgo~190425z and Ligo/Virgo~190510g 
at redshift $z \sim 0.06$ and $z \sim 0.05$, respectively. For latter events, which were about 6 
times further than GW/GRB~170817A only upper limits on any electromagnetic emission is available. 
Additionally, the host of the faint X-ray dark short GRB~080121~\cite{grb080121} may belong to a 
galaxy group at redshift $z = 0.046$~\cite{grb080121redshift}. If true, this burst was intrinsically 
only 2 times brighter than GRB~170817A, which its $E_{iso}$ was a few orders of magnitude fainter 
than typical short bursts, see Fig. \ref{fig:allsgrb}. 

The faintness or absence of GRB in gravitation wave events associated to BNS may be interpreted 
as a natural consequence of jets directionality. However, it does not explain the absence of 
optical emission from ejected disk/torus, which its emission is roughly spherical - presumably due 
to its faintness. Therefore, despite the impact of off-axis view of GW/GRB~170817A, intrinsic 
properties of the progenitor BNS, in particular magnetic field of neutron stars, their 
spin, and their age, might have been involved in its weakness. In~\cite{hourigw170817} this subject 
is discussed in detail and we summarize its conclusions in Sec. \ref{sec:progenitor}.

\subsubsection{Delayed brightening}
In the literature, along with its intrinsic faintness, the late brightening of GW/GRB~170817A 
afterglows is usually considered as an exceptional characteristic of this bursts. 

Afterglows of GRBs up to few thousands of seconds are most probably a superposition of weak internal 
shocks in what remains from the relativistic jet after the main prompt shock, and emission from 
external shocks~\cite{xrtafterglow}. Assuming a narrow prompt spike, the delay between the prompt 
and onset of afterglow emission is $\Delta t \sim r_e / 2c \Gamma^2$, where $c$ is the speed of 
light. For the model of Table \ref{tab:agparam} $\Delta t \sim 4$~sec. This delay is much shorter 
than any automatic follow up and practically unobservable. Thus, the shape of X-ray light curve 
depends on relative importance of decreasing emission from internal shocks and increasing emission 
from external shocks. The absence of early brightening or even a plateau in many short GRBs means 
that their early X-ray is dominated by what we call {\it tail emission} of internal shocks. Thus, 
as pointed out earlier, in absence of long follow up observation of short GRBs we do not usually 
detect the second component and ignore how its light curve look like at late times. Short GRB's 
with a plateau in their X-ray light curve may be those in which the afterglow takes off quickly 
and peaks a few days after prompt gamma-ray.

The initial suggestions about brightening of afterglows due to the reduction of off-axis view when 
the jet become dissipated and its content scatter to our line of sight, is not consistent with 
relatively early break of light curves in all three energy bands after $\sim T+110$~days. Therefore, 
viewing angle cannot be the only reason for the late brightening of afterglows. On the other hand, 
our simulations show that in the case of GW/GRB170817A the slow rise of the afterglow was due to 
the long distance of ISM/circum-burst material from center, its low density and low column density 
of the jet. Dilution of the jet was partially due to the intrinsically low density and low Lorentz 
factor of polar ejecta - at least in our direction - and partially the result of large distance of 
surrounding material from center, that is $\sim 1000$~AU (see Table \ref{tab:agparam}) rather than 
e.g. $\sim 200$~AU for recently detected NIR emitting material around the isolated neutron star 
RXJ0806.4-4123~\cite{nstarmatt}. Consequently, jet's column density was extensively reduced by 
lateral expansion.

It is however cautious to consider that our conclusion about the effect of distance may be somehow 
biased and correlated to spatial resolution of our simulations. Nonetheless, estimation of shock 
distance for this GRB in the literature and for other GRBs with various methods, such as cooling of 
thermal emission~\cite{grb100316d} show that the range of distances obtained from simulation of 
GRBs by the formalism used here~\cite{hourigrbmag} are realistic.

\section {Properties of GW/GRB~170817 progenitors and their environment} \label{sec:progenitor}
If the unusual properties of GW/GRB~170817A and presumed faintness of other nearby candidate BNS 
mergers are merely due to an off-axis view, they tell little about properties of their progenitor 
neutron stars. By contrast, if the faintness is at least in some extent intrinsic, it may 
be an evidence for evolution of BNS population. Assuming an intrinsic origin, in this section we 
use results of GRMHD simulations of BNS merger and observed properties of neutron stars and 
pulsar, which present a relatively young population with an age less than $\mathcal{O}(1)$~Myr, 
to understand properties of the nearby, i.e. $z \lesssim 0.1$ BNS and their merger.

\subsection {Equation of state, magnetic field and spin} \label{sec:proghist}
In addition to the relation between mass and radius, the Equation of State (EoS) of neutron stars 
determines core and crust density and buoyancy, and thereby their tidal 
deformability~\cite{nstarstaterev} which affects high frequency ringdown of gravitational waves 
at the end of binary's inspiral. Although resolution of LIGO-Virgo at high frequency is not 
sufficient for detailed discrimination between equations of states, it is sufficient for 
distinguishing between stiff, that is high pressure, and soft EoS. In the case of GW~170817 stiff 
equations of state are disfavored~\cite{gw170817ligo}. For close mass NS progenitors - as it was 
the case of GW~170817 - the density of inner part of the accretion disk and poloidal magnetic 
field of the merger are smaller in soft EoS~\cite{nsmergerrprocsimul0,diskjetcorrel,blandfordzenajek}.

Simulations of~\cite{nsmergerrprocsimul0} show that in equal mass BNS merger, if the initial magnetic 
fields of progenitors are aligned with each other and anti-aligned with the rotation axis of the 
BNS, average poloidal magnetic field of the polar ejecta is about 5 times weaker than when both 
initial fields are aligned with the rotation axis. Additionally, if the initial fields are 
anti-aligned with each other, the average poloidal field is even smaller by a factor of few. 
Anti-alignment may happen through interaction of magnetized material surrounding progenitors before 
the last stages of inspiral. In this case, reduction of the field further from rotation axis is 
larger. These characteristics have direct impact on attainable Lorentz factor when ejected polar 
material is accelerated by transfer of magnetic to kinetic energy~\cite{grbjetsimul}. On the other 
hand, star population of the host galaxy NGC 4993 is relatively 
old~\cite{gw170817ligohost,gw170817ligohost0,gw170817ligohost1}. Therefore, magnetic fields of 
the progenitor neutron stars of GW~170817 might have been as low as $10^8 - 10^9$ G and fast 
precessing, if the progenitors were recycled millisecond pulsars or even smaller if they had 
evolved in isolation~\cite{nstarrev,nstarrev0}. 

Initial spins of progenitor neutron stars, which carry a fingerprint of their evolution history, 
have a crucial role in the dynamics of their merging. In particular, they affect the amount of 
ejecta, density and extent of accretion disk/torus, and spin of the short living High Mass Neutron 
Star (HMNS) and the final black hole. Unfortunately, GW~170817 event was too faint and spin of 
progenitors could not be determined from gravitational waves.
 
When spin axes are aligned with orbital rotation, binding energy of BNS is weaker, inspiral regime 
is longer and amount of ejected mass is larger too. However, the latter depends on the mass ratio of 
progenitors and is smaller for equal mass binaries~\cite{nstarmergespin}. It is clear that direction 
of these differences are similar to those of magnetic field. However, spin effect on the amount of 
ejected material is subdominant with respect to other processes and can modify it by only a few 
percents~\cite{nstarmergespin}. 

In summary, the observed properties of GW and electromagnetic emissions of GW~170817 event are 
consistent with each others and point to a reduced ejecta and magnetic field, and thereby a weak GRB.

\subsection {Environment of progenitor BNS} \label{sec:progenv}
The afterglow model described in Sec. \ref{sec:ag} requires that density of external material 
depend on the azimuthal angle with respect to jet axis. Despite degeneracy of the parameters, 
the directional anisotropy of ISM/circum-burst material may be real because we could not find 
consistent model without it. This means that circum-burst material was not only the ISM. Moreover, 
origin of additional material and its properties were somehow correlated with progenitors. 

In young neutron stars and pulsars the distance to wind Termination Shock (TS) $R_{TS} $ depends on 
the rate of mass loss and wind pressure inside wind nebular, which is balanced by the ISM pressure, 
and its typical value is $R_{TS} \sim \mathcal{O}(0.1)$~pc~$\sim \mathcal{O}(10^{17})$~cm
~\cite{nssheath}. In old neutrons it is expected that the reduction of glitching activities due 
to the solidification of crust and dissipation of magnetic field gradually decreases mass 
loss and $R_{TS}$. However, detection of this population is very difficult and information about 
their properties is extremely rare. An exception is the isolated neutron star RXJ0806.4-4123 
with an age of $\sim 10$~Myr for which thermal material at a relatively short distance of 
$\sim 200$~AU$~\sim 3 \times 10^{15}$~cm is detected~\cite{nstarmatt}. 

Our simulations of the afterglows estimates $R_{TS} \sim 10^{16}$~cm, which is between the above 
values. The reason may be an age larger than most observed neutron stars and pulsars, estimated to be 
$\lesssim \mathcal{O}(10^4)$~yr, and younger than $\sim 10$~Myr age of RXJ0806.4-4123. It is also 
plausible that the progenitor neutron stars were even older, but during early stages of inspiral, 
that is well before generation of detectable gravitational waves, strong tidal forces had induced 
crustal faults and resumed glitching and mass loss. If this explanation is correct, ejected material 
had to also persist at shorter distances and inside the wind bubble. To estimate their average 
column density we employ distribution used in the afterglow model for ISM/circum-burst material, 
that is $N'(r) = N'(r_0) (r/r_0)^{-\kappa}$ and extend it to $r < r_0$. For model C1 this estimation 
gives a column density of $\sim 4 \times 10^{14}$~cm$^{-2}$, which is much smaller than column density 
of the components of model. It is also smaller than swept material in the first 
$\sim 3 \times 10^5$~sec after the onset of the external shocks, and therefore completely negligible. 
On the other hand, if we consider much denser circum-burst material at shorter distances, much higher 
X-ray flux generated by the shocks violates upper limits at $\sim T+2$~days, see additional 
simulations in~\cite{hourigw170817lateagjet}. Using these two extreme cases, we conclude that the 
column density of material inside $r_e \sim 10^{16}$~cm bubble was 
$< \mathcal{O} (1)\times 10^{15}$~cm$^{-2}$ or equivalently its average density was $< 0.4$~cm$^{-3}$.

\section{Outline and prospectives for future} \label{sec:prespect}
In summary, from analysis of prompt and afterglows of GW/GRB170817A along with information acquired 
from gravitational wave observation we can draw the following consistent picture for the first BNS 
merger event extensively followed up in multi-wavelength by multi-probe instruments:

\begin{description}
\item{-} Progenitors were old and cool neutron stars with close masses; 
\item{-} They had soft equations of state and small initial magnetic fields of $\lesssim 10^9$ G. 
Their fields were anti-aligned with respect to orbital rotation axis and each other.
\item{-} For dynamical and/or historical reasons such as encounter with similar mass objects, their 
spins before final inspiral were anti-aligned.
\item{-} The merger produced a HMNS with a moderate magnetic field of $\lesssim 10^{10}$ G. 
This value is in the lowest limit of what is obtained in typical GRMHD simulations. 
\item{-} The HMNS eventually collapsed to a black hole and created a moderately magnetized 
disk/torus and a low density, low magnetized outflow.
\item{-} A total amount of $\sim 0.03-0.05~M_\odot$ material, including $10^{-3} - 10^{-2}~M_\odot$ of 
tidally stripped pre-merger and a post-merger wind were ejected to high latitudes. They were 
subsequently collimated and accelerated by transfer of Poynting to kinetic energy. The same 
process increased electron yield by segregation of charged particles.
\item{-} A small mass fraction of the polar ejecta was accelerated to ultra-relativistic velocities 
and made a relatively weak GRB. The reason for low Lorentz factor, density, and extent of this 
component was the weakness of the magnetic field. 
\item{-} For the same reasons, the ultra-relativistic section of the jet was narrow and our off-axis 
view of $\sim 10^\circ$ was enough to reduce the emission of high energy photons in our direction.
\item{-} After prompt internal shocks the jet had a wide angular distribution with varying density 
and Lorentz factor. But despite significant energy dissipation in its core, a tiny fraction of 
the jet had preserved its coherence and boost up to its collision with circum-burst material at 
$\sim 10^{16}$ cm from merger.
\item{-} In addition to the ISM, circum-burst material included a component which its origin was 
correlated with the BNS.
\item{-} The late brightening of afterglows was due to the relatively long distance of circum-burst 
material and low density of material inside the wind bubble surrounding the BNS.
\item{-} Angular variation in the jet was also responsible for domination of emission in lower 
energies from high latitude side lobes and thereby observation of a superluminal motion of the 
radio afterglow.
\end{description}

This qualitative picture and redshift distribution of short GRBs point to an evolutionary effect 
on their properties. This conclusion is somehow strengthen by failure of finding an electromagnetic 
counterpart for two recent candidate BNS merger gravitational wave events. Although the absence of a 
GRB may be due to our off-axis viewing angle, it does not explain the failure of kilonova detection. 
In any case, number of such events is still too small to make any statistically meaningful 
conclusion. Nonetheless, this analysis demonstrates the importance of multi-probe observation of 
this and other categories of transients detected through their gravitational waves. 

A highly desired capability is to detect compact object collisions well before their merger such 
that the evolution of the source can be followed by other probes. This needs gravitational wave 
detectors working at lower frequencies, such as LISA. On the other hand, higher frequency bands 
are necessary for studying ringdown regime. They enable us to follow deformation of merging objects, 
which can be then related to a state of quantum matter and QCD interaction impossible to reproduce 
in laboratory. They are also important for testing gravity and quantum gravity models, for instance 
by detecting gravitational wave echos generated by quantum process, 
see e.g.~\cite{gwbhecho,gwbhecho0}. Additionally, improving angular resolution of GW detectors is 
crucial for accelerating follow up by other probes and increasing the chance of detecting 
electromagnetic counterparts. Synchronized and automatic wide field observations in UV/optical/IR are 
important because apriori they do not have the issue of directionality of high energy emission from 
a narrow jet.

Gravitational wave detectors have opened a whole new channel by which we can {\it see} events and 
phenomena completely hidden from us until now. Nonetheless, complementary observations of 
electromagnetic emission, neutrinos and cosmic rays from these events are necessary for achieving a 
full understanding of underlying physical processes.

\appendix

\section{Definition of parameters and models of active region} \label{app:paramdef}
Table \ref{tab:paramdef} summarizes parameters of this model. Despite their long list, simulations 
of typical long and short GRBs in~\cite{hourigrbmag} show that the range of values which lead to 
realistic bursts are fairly restricted. 

\begin{table}[H]
\begin{center}
\caption{Parameters of the phenomenological prompt model \label{tab:paramdef}}
\vspace{0.5cm}
\begin{tabular}{|p{2.5cm}|p{10cm}|}
\hline
~Model (mod.) & Model for evolution of active region with distance from central engine; See 
Appendix \ref{app:paramdef} and~\cite{hourigrb,hourigrbmag} for more details. \\
~$r_0$ (cm) & Initial distance of shock front from central engine. \\
~$\Delta r_0$ & Initial (or final, depending on the model) thickness of active region.~ \\
~$p$ & Slope of power-law spectrum for accelerated electrons; See eq. (3.8) of~\cite{hourigrbmag}.~ \\
~$p_1,~p_2$ & Slopes of double power-law spectrum for accelerated electrons; See eq. (3.14) 
of~\cite{hourigrbmag}. \\
~$\gamma_{cut}$ & Cut-off Lorentz factor in power-law with exponential cutoff spectrum for 
accelerated electrons; See eq. (3.11) of~\cite{hourigrbmag}. \\
~$\gamma'_0$ & Initial Lorentz factor of fast shell with respect to slow shell. \\
~$\tau$ & Index in the model defined in eq. (3.28) of~\cite{hourigrbmag}. \\
~$\delta$ & Index in the model defined in eq. (3.29) of~\cite{hourigrbmag}. \\
~$Y_e$ & Electron yield defined as the ratio of electron (or proton) number density to baryon 
number density. \\
~$\epsilon_e$ & Fraction of the kinetic energy of falling baryons of fast shell 
transferred to leptons in the slow shell (defined in the slow shell frame). \\
~$\alpha_e$ & Power index of $\epsilon_e$ as a function of $r$. \\
~$\epsilon_B$ & Fraction of baryons kinetic energy transferred to induced magnetic field in 
the active region. \\
~$\alpha_B$ & Power index of $\epsilon_B$ as a function of $r$. \\
~$N'$ & Baryon number density of slow shell. \\
~$\kappa$ & Power-law index for N' dependence on $r'$. \\
~$n'_c$ & Column density of fast shell at $r'_0$. \\
~$\Gamma^\dagger$ & Lorentz factor of slow shell with respect to far observer. \\
~$|B|$ & Magnetic flux at $r_0$. \\
~$f$ & Precession frequency of external field with respect to the jet. \\ 
~$\alpha_x$ & Power-law index of external magnetic field as a function of $r$. \\
~$\phi$ & Initial phase of precession, see ~\cite{hourigrbmag} for full description. \\
\hline
\end{tabular}
\end{center}
{\small
\begin{description}
\item{$\star$} The phenomenological model discussed in~\cite{hourigrb} and its 
simulation~\cite{hourigrbmag} depends only on the combination $Y_e\epsilon_e$. For this reason 
only the value of this combination is given for simulations.
\item{$\star$} The model neglects variation of physical properties along the jet or active region. 
They only depend on the average distance from center $r$, that is $r-r_0 \propto t-t_0$.
\item{$\star$} Quantities with prime are defined with respect to rest frame of slow shell, and 
without prime with respect to central object, which is assumed to be at rest with respect to 
a far observer. Power indices do not follow this rule.
\item{$\dagger$} According to this definition for external shocks $\Gamma_e \approx 1$ (index $e$ 
for {\it external shock})and $\gamma'_0 = \Gamma_i \equiv \Gamma$ (index $i$ for {\it internal shock}. 
\end{description}
}
\end{table}

In the phenomenological model of~\cite{hourigrb} the evolution of $\Delta r'(r')$ cannot be 
determined from first principles. For this reason we consider the following phenomenological 
models:
\bea
&& \Delta r' = \Delta r'_0 \biggl (\frac {\gamma'_0 \beta'}{\beta'_0 \gamma'} 
\biggr )^{\tau}\Theta (r'-r'_0) \quad \text {dynamical model, Model = 0} \label {drdyn} \\
&& \Delta r' = \Delta r'_{\infty} \bigg [1-\biggl (\frac{r'}{r'_0} \biggr )^
{-\delta}\biggr ] \Theta (r'-r'_0) \quad \text {Steady state model, Model = 1} \label {drquasi} \\
&& \Delta r' = \Delta r'_0 \biggl (\frac{r'}{r'_0} \biggr )^{-\delta} 
\Theta (r'-r'_0) \quad \text {Power-law model, Model = 2} \label {drquasiend} \\
&& \Delta r' = \Delta r_{\infty} \bigg [1- \exp (- \frac{\delta(r'-r'_0)}{r'_0}) \biggr ] 
\Theta (r-r'_0) \quad \text {Exponential model, Model = 3} \label {expon} \\
&& \Delta r' = \Delta r'_0 \exp \biggl (-\delta\frac{r'}{r'_0} \biggr )
\Theta (r'-r'_0) \quad \text {Exponential decay model, Model = 4} \label {expodecay}
\eea
The initial width $\Delta r'(r'_0)$ in Model = 1 \& 3 is zero. Therefore, they are suitable for 
description of initial formation of an active region in internal or external shocks. Other models 
are suitable for describing more moderate growth or decline of the active region. In tables of models 
the column $mod.$ corresponds to numbers given in \ref{drdyn}-\ref{expodecay} and indicates which 
evolution rule is used in a simulation regime.

\end{document}